\documentclass[
 reprint,
superscriptaddress,
 amsmath,amssymb,
 aps,
prb,
floatfix,
]{revtex4-2}

\usepackage{graphicx}
\usepackage{dcolumn}
\usepackage{bm}
\usepackage{braket}
\usepackage{svg}
\usepackage{xcolor}
\usepackage[T1]{fontenc}

\renewcommand{\vec}[1]{\mathbf{ #1}}

\begin{document}

\preprint{APS/123-QED}

\title{Carrier Mobility of Strongly Anharmonic Materials from First Principles}

\author{Jingkai Quan}
\affiliation{
 The NOMAD Laboratory at Fritz-Haber-Institut der Max-Planck-Gesellschaft, \\ Faradayweg 4-6, 14195, Berlin, Germany
}%
\affiliation{
 Max-Planck Institute for the Structure and Dynamics of Matter, Luruper Chausse 149, 22761, Hamburg, Germany
 }%

\author{Christian Carbogno}%
\affiliation{
 The NOMAD Laboratory at Fritz-Haber-Institut der Max-Planck-Gesellschaft, \\ Faradayweg 4-6, 14195, Berlin, Germany
}%

\author{Matthias Scheffler}

\affiliation{
 The NOMAD Laboratory at Fritz-Haber-Institut der Max-Planck-Gesellschaft, \\ Faradayweg 4-6, 14195, Berlin, Germany
}%

\date{\today}

\begin{abstract}
First-principle approaches for phonon-limited electronic transport are typically based on many-body perturbation theory and transport equations.
With that, they rely on the validity of the quasi-particle picture for electrons and phonons, which is known to fail in strongly anharmonic systems. 
In this work, we demonstrated the relevance of effects beyond the quasi-particle picture by combining {\it ab initio} molecular dynamics and the 
Kubo-Greenwood (KG) formalism to establish a non-perturbative, stochastic method to calculate carrier mobilities while accounting 
for all orders of {anharmonic and electron-vibrational} couplings. In particular,
we propose and exploit several numerical strategies that overcome the notoriously slow convergence of the KG formalism for both electronic and nuclear 
degree of freedom in crystalline solids.  
The capability of this method is demonstrated by calculating the temperature-dependent electron mobility of the strongly anharmonic oxide perovskites SrTiO$_3$ 
and BaTiO$_3$ across a wide range of temperatures. We show that the temperature-dependence of the mobility is largely driven by anharmonic, higher-order
coupling effects and rationalize these trends in terms of the non-perturbative electronic spectral functions. 
\end{abstract}

\maketitle

\section{Introduction}
Charge transport plays {an important} role in the field of condensed matter physics, material science, and engineering since over hundred years~\cite{drude1900elektronentheorie,ashcroft2022solid}. 
In fact, this topic is of pivotal importance for technological applications, since it determines the performance of {batteries~\cite{battery_review}}, transistors~\cite{transistors_perspective}, 
thermoelectric elements~\cite{thermoelectric_review,thermoelectric_conductivity_example}, solar cells~\cite{bernardi_si}, and more. At the macroscopic level, charge transport and the carrier conductivity~$\bm{\sigma}$ is defined by 
Ohm's law that relates the applied electric field ${\textbf E}(t)$ and the induced current~${\textbf J}(t)$:
\begin{equation}
    \label{Ohm_law}
    {\textbf J}(t) =\bm{ \sigma } {\textbf E}(t) \;,
\end{equation}
where the conductivity $\bm{\sigma}$ is a $3\times3$ tensor. {In general,} there are different mechanisms associated with charge transport in solids: band-like conductivity originating from Bloch-type quasiparticles~\cite{ponce_review}, 
hopping conductivity from polaron motion~\cite{polaron_review}, and ionic conductivity associated with the diffusion of ions~\cite{fast_ionic_conductor}. In all of these cases, two distinct 
quantities drive the respective conductivity,~i.e.,~the amount of mobile, ``free'' charge carriers~$n$~(electrons, holes, polarons, anions, or cations) and the mobility~$\mu$ of these charge carriers. The conductivity can thus be expressed as
\begin{equation}
\bm{\sigma} = {qn}\bm{\mu} \;,
\label{eq:conductivity_defination}
\end{equation}
in which $q$ is the charge that the respective particle carries. In this spirit, it is common to measure charge carrier concentrations and mobility independently and to report the mobility as a function
of the charge carrier density~\cite{ponce_review}.

In this work, we focus on band-like conductivity, the dominant mechanism in semiconductors~\cite{ashcroft2022solid}. In this regard, decades of research in band-structure engineering and doping
have focused on understanding and tailoring the band structure and charge carrier density for specific applications~\cite{band_engineering_review_npj, band_engineering_dopping, band_engineering_nc,band_engineering_solar}. In contrast, less is known about the band-like mobility itself, which is limited by different scattering mechanisms. For instance, electron-electron scattering plays an important 
role at high electron densities,~e.g.,~in metals~\cite{e-e_interaction_BTE}, while electron-defect scattering~\cite{I-te_prm} is known to be the limiting factor at very low 
temperatures~\cite{I-te_prm,I-te_prm_2022}. For non-degenerate semiconductors at ambient temperature and above, the mobilities are essentially determined by 
electron-phonon scattering,~i.e.,~the interaction of the electronic charge carriers with the thermal vibrations of the nuclei~\cite{ponce_review}. 
To address the latter effect from first-principles, it is common to combine density-functional theory~(DFT)~and many-body perturbation theory~(MBPT)~\cite{giustino_rmp}. In this framework, 
the nuclear dynamics is approximated in terms of phonons,~i.e.,~one assumes that the nuclei move on a single, perfectly harmonic potential-energy valley. 
Similarly, it is presumed that the associated electron-phonon scattering is weak enough to be approximated in terms of the first-order response to nuclear displacements,~i.e.,~the so called electron-phonon coupling elements. In turn, MBPT then provides the (vibrationally) renormalized {electronic} self-energy. 
Its real-part describes the band structure renormalization,~i.e.,~energy shifts in the electronic states due to e-ph scattering, whereas the imaginary part corresponds to the el-ph scattering cross-sections. In conjunction with the Boltzmann transport equation~(BTE) and/or the Wigner transport equation~(WTE)~\cite{wte_conductivity}, this allows to compute phonon-limited mobilities from first principles~\cite{bernardi_si,bte_diamond_prb,wuli_bte,ponce_prr,ponce_review,phoebe,epw,perturbo}.  

However, the mentioned approximations on which these transport equations rely may well fail, especially in complex materials and/or at elevated temperatures.
As show recently in the closely related field of vibrational heat transport, so called anharmonic effects beyond the harmonic 
approximation can play a decisive role in the nuclear dynamics even at room temperature~\cite{florain_prm}. 
When such anharmonic effects {are} too strong, the phonon picture itself becomes invalid, as the phonon lifetime may get comparable or even shorter than the period of just one phonon oscillation
when violating the Ioffe-Regel limit~\cite{ioffe1960non,simoncelli_prx}. 
{Then}, also the BTE and WTE become no longer applicable, since they rely on the validity of the phonon picture. For example, it has been shown that this case
occurs when  short-lived intrinsic defects are formed in crystalline solids~\cite{florian_gk_application}. Such finite-temperature effect may resemble 
local structures of other crystal phases, but they may occur already at temperatures well below the relevant phase transition.
In such cases, accurate transport predictions require methods that explicitly take into accounts all orders of {anharmonicity}~\cite{chris_gk,florian_gk_implementation} {and electron-vibrational coupling}.
For instance, this has been also observed in oxide (SrTiO$_3$) and halide (CsPbBr$_3$) perovskites~\cite{marios_bzu,marios_perovskites_npj}, for which the real part of the
electronic self-energy is substantially influenced by higher-order couplings. This suggests that these effects might also substantially affect mobility predictions.

Several different methods have been developed {to account (at least partially) for anharmonic and/or higher-order coupling effects~\cite{sscha,tdep_olle_prb,marios_special_dispalcement}. 
For example, renormalized, effectively temperature-dependent phonon theories had been proven useful in practice, 
resulting in improved predictions~\cite{thermal_conductivity_spectral_function_method,sto_bte}. However, also these approaches inherently rely on the validity of 
the phonon picture and are hence inapplicable beyond that regime~\cite{simoncelli_prx}.
To address this regime, a non-perturbative theory of charge transport in crystalline materials is needed. 

{A} promising route to overcome the described limitations of MBPT in the prediction of mobilities is the Kubo-Greenwood (KG) formalism~\cite{Greenwood_original}.
In this approach, all anharmonic effects can be taken into account via {\it ab initio} molecular
dynamics~({\it ai}MD) simulations. Similarly, all orders of vibrational coupling are accounted for by explicitly considering the 
evolution of the electronic-structure during the dynamics within the Born-Oppenheimer approximation. Let us emphasize that this naturally includes
the aforementioned intrinsic defect formation~\cite{florian_gk_application}, both at the structural level during the {\it ai}MD dynamics and at the electronic level, since
the electronic-structure at each step is evaluated self-consistently. 
For disordered systems, {the KG approach} has been successfully applied to 
determine conductivities,~e.g.~for warm dense plasmas\cite{holst_prb, plasma_pre_2017, plasma_pre_2022}, 
liquids~\cite{vasp_kg_2002,vasp_al_test_2013, liquid_al_prb_2005}, and amorphous compounds~\cite{abtew_dc_limit, cu3sbs4_kg}.
For ordered, crystalline systems, the application of the KG formalism is, however, {technically challenging}~\cite{qe_kg_cpc,plasma_pre_2022,vasp_al_test_2013,Zhenkun_2022}. 
The fundamental reason is that crystalline systems are typically characterized by dispersive electronic and vibrational states. For the KG approach, this results in a substantial computational effort: 
On one hand, large supercells are required to account for long-wavelength lattice vibrations in the {\it ai}MD simulations; on the other hand, dense reciprocal-space {\bf k}-grids are needed for the Brillouin zone integrations to {properly account for} the dispersive character of the electronic structure. So far, these computation limitations have prevented a fully anharmonic assessment of mobilities in crystalline semiconductors~\cite{Zhenkun_2022}.

In this work, we discuss and demonstrate several strategies to overcome the numerical challenges described above, which eventually allows to account for all orders of electron-vibrational and anharmonic couplings when predicting mobilities. To this end, we introduce the theoretical fundamentals of the Kubo-Greenwood formalism and the resulting first-principles workflow in Sec.~\ref{sec:theory_kg}.  In Sec.~\ref{sec:implementation_kg}, we discuss the 
details of our implementation in the all-electron, numeric atom-centered orbitals (NAOs) based {\it ab initio} materials simulation package \texttt{ FHI-aims}~\cite{fhi-aims}. In particular, we analyze the strategies used to tackle the numerical challenges mentioned above. In Sec.~\ref{sec: application_kg}, we then demonstrate the merits of the method by calculating temperature-dependent electron mobilities for the strongly anharmonic perovskites SrTiO$_3$ 
and BaTiO$_3$ up to high temperatures. The results are rationalized by analyzing the associated, temperature-dependent spectral functions. Finally, in Sec.\ref{sec:conclusion_kg}, we summarize the 
results and discuss future opportunities to improve the {efficiency} of the {\it ab initio} KG formalism.

\section{The Kubo-Greenwood Formalism}
\label{sec:theory_kg}

In linear response theory, the {carrier conductivity} $\bm{\sigma}(t)$ describes the linear proportionality between an applied, time-dependent electromagnetic field $\textbf{E}(t)$ and the resulting current $\textbf{J}(t)$, in close analogy to Ohm's law in Eq.~(\ref{Ohm_law}). 
Fourier transformation with respect to time yields the frequency-dependent conductivity $\bm{\sigma}(\omega)$, a quantity that is often referred to as optical conductivity~\cite{ponce_review}.
Here, the frequency~$\omega$ describes the applied alternating-current~(AC) field; its static limit ($\omega\to 0$) is the direct current (DC) conductivity.} 

The Kubo formalism~\cite{kubo_original} express the frequency-dependent conductivity tensor at finite temperature in terms of current-current correlation functions~\cite{holst_prb}:
\begin{equation} \label{eq:kubo-formula}
    \sigma_{ij}(\omega) = \lim_{\alpha \to 0^+} \int^{\infty}_{0} dt e^{i(\omega+i\alpha)t} \int^{\beta}_0 d\tau \textbf{Tr}[\hat{\rho}_0\hat{\textbf{J}}_i (t-i\hbar\tau)\cdot\hat{\textbf{J}}_j] \;.
\end{equation}
Here $i,j \in \{x,y,z\}$ denote the Cartesian axes, $\beta = (k_B T)^{-1}$, {and $\alpha \to 0^+$ is an infinitesimal adiabatic turn-off parameter for the response. 
Physically, it reflects the causal relation between drive and response~\cite{baroni_kubo_arrow_of_time} and hence ensures 
that the response function vanishes at large times~\cite{bruus_many_body}, when the system is back in equilibrium. In turn,
this guarantees the mathematical converge of the integral in Eq.~(\ref{eq:kubo-formula}).}

{Formally, the general Kubo formula in Eq.~(\ref{eq:kubo-formula}) covers all kind of charge transport mechanisms, including polaronic~\cite{kubo_polaron_conductivity} and ionic~\cite{kubo_ionic_conductivity} conductivity, as long as the respective current operator accounts for the respective transport process. In this work, 
we focus on band-like conductivity and hence employ the respective, well-known expression for the band-like current operator in terms of effective, single-particle Bloch states $\ket{\textbf{k}\mu}$~\cite{Greenwood_original,holst_prb}:} 
\begin{equation}
\label{eq: IPA}
    \hat{\textbf{J}} = \frac{q}{m}\sum_{\textbf{k}\mu\nu} \braket{\textbf{k}\mu|\hat{\textbf{p}}|\textbf{k}\nu}\ket{\textbf{k}\mu}\bra{\textbf{k}\nu} \;.
\end{equation}
Here, $\textbf{k}$ is the wave vector and $\mu$ is the band index of the Bloch state, $\hat{\textbf{p}} = m\hat{\textbf{v}} = i\hbar\nabla$ denotes the momentum operator, 
and $\hat{\textbf{v}}$ indicates the velocity operator. Let us emphasize that the effective, single-particle Bloch states $\ket{\textbf{k}\mu}$ entering Eq.~(\ref{eq: IPA}) are expected to correctly 
describe the quasi-particle excitation spectrum. Accordingly, they should be computed with an appropriate many-body formalism,~e.g.,~$GW$~\cite{gw_review}. {In the present work, we will approximate 
them by Kohn-Sham states obtained from DFT, though.} The limitations of this choice are discussed in more detail in Sec.~\ref{def_mobility1}.

Within this effective single-particle approximation, Eq.~(\ref{eq:kubo-formula}) yields the following Kubo-Greenwood (KG) formula for the conductivity~\cite{Greenwood_original}:
\begin{eqnarray} \label{eq:complex_kg}
    \sigma_{ij}(\omega) = \lim_{\alpha \to 0^+} && i\frac{2q^2\hbar^3}{m^2V}\sum_{\textbf{k}\mu\nu} \frac{(f_{\textbf{k}\nu}-f_{\textbf{k}\mu})}{(\epsilon_{\textbf{k}\mu}-\epsilon_{\textbf{k}\nu})} \nonumber\\
    && \times \frac{\braket{\textbf{k}\nu|\nabla_i|\textbf{k}\mu}\braket{\textbf{k}\mu|\nabla_j|\textbf{k}\nu}}{\epsilon_{\textbf{k}\mu}-\epsilon_{\textbf{k}\nu}-\hbar\omega + i\alpha} \;,
\end{eqnarray}
where $\epsilon_{\textbf{k}\mu}$ and $f_{\textbf{k}\mu}$ are the energy eigenvalue and the occupation number of state $\ket{\textbf{k}\mu}$; the summation 
is over all $\textbf{k}$ points in the first Brillouin zone of the utilized cell. 
Note that the resulting conductivity tensor ${\bm \sigma}(\omega)$ is complex; its real and imaginary part are related by the Kramers-Krönig relation, in close analogy to the permittivity
defined in classical electrodynamics~\cite{ashcroft2022solid,jackson_ed}. Similarly, the imaginary part of ${\bm \sigma}(\omega)$ describes if the response is 
capacitative or inductive, whereas the real part denotes the actual conductivity of interest in this work. Accordingly, we focus on the real part for the remainder of this paper. 

Eventually, we apply the Born-Oppenheimer approximation,~i.e.,~we assume that the electrons follow the nuclei instantaneously, and hence 
take the limit $\alpha \to 0^+$ of instantaneous response. The obtained off-diagonal elements ($i \neq j$) of ${\bm \sigma}(\omega)$ describes the intrinsic anomalous Hall effect in the DC limit~\cite{ahe_rmp,ahe_wannier_inter}:
\begin{eqnarray}  \label{eq:ahe_kubo}
    \Re(\sigma_{ij}(\omega=0)) =&& \frac{2\pi q^2\hbar^3}{m^2V}\sum_{\textbf{k}\mu\nu} 
    \Im(\braket{\textbf{k}\nu|\nabla_i|\textbf{k}\mu}\braket{\textbf{k}\mu|\nabla_j|\textbf{k}\nu}) \nonumber\\
    && \times \frac{(f_{\textbf{k}\nu}-f_{\textbf{k}\mu})}{(\epsilon_{\textbf{k}\nu}-\epsilon_{\textbf{k}\mu})^2} \;,
\end{eqnarray}
and the diagonal elements ($i = j$) describes the desired longitudinal conductivity:
\begin{eqnarray}  \label{eq:kg-formula}
    \Re(\sigma_{ii}(\omega)) =&& \frac{2\pi q^2\hbar^2}{m^2V\omega}\sum_{\textbf{k}\mu\nu} 
    (\braket{\textbf{k}\nu|\nabla_i|\textbf{k}\mu}\braket{\textbf{k}\mu|\nabla_i|\textbf{k}\nu}) \nonumber\\
    && \times (f_{\textbf{k}\nu}-f_{\textbf{k}\mu}) \delta(\epsilon_{\textbf{k}\mu}-\epsilon_{\textbf{k}\nu}-\hbar\omega) \;.
\end{eqnarray}
Due to its similarity to Fermi's golden rule, Eq.~(\ref{eq:kg-formula}) can be rationalized in terms of instantaneous scattering events
between different electronic states $\ket{\textbf{k}\mu}$ and $\ket{\textbf{k}\nu}$ with the same crystal momentum~$\textbf{k}$~(vertical transition).
Accordingly, the scattering amplitude is given by the momentum matrix element $i\hbar\braket{\textbf{k}\mu|\nabla|\textbf{k}\nu}$ while the delta function ensures energy conservation. 
At first glance, indirect  transitions,~i.e.,~transitions between electronic states with different crystal momentum~$\vec{k}$, appear not to 
enter the KG formula. However, one has to keep in mind that the KG formula needs to be evaluated in extended 
supercells. For the electronic degrees of freedom, this results in a reduction of the first Brillouin zone~(BZ), 
often called ``Brillouin zone folding''. In turn, indirect transitions in the primitive BZ formally become 
vertical transitions in the folded BZ. By this means, they are effectively accounted for in the KG formula 
in Eq.~(\ref{eq:kg-formula}), as discussed in more detail in Sec.~\ref{sec:extrapolation}.

Eventually, let us note that Eq.~(\ref{eq:kg-formula}) denotes the conductivity for one specific nuclear configuration.
In order to obtain conductivities at finite temperature $T$, it is thus necessary 
to sample the available phase space in the canonical ensemble and perform an ensemble 
average~$\braket{\sigma(\omega)}_T$. In practice, this implies evaluating the average of the 
KG formula over many different nuclear configuration, since those configurations enter the 
KG only indirectly via the Kohn-Sham Hamiltonian, its eigenvalues, and eigenstates~$\ket{\textbf{k}\mu}$. 
For crystalline systems, this means that extended supercells need to be used, so that also 
vibrational degrees of freedom with long wavelengths are appropriately sampled, which in turn 
ensures that indirect electronic transitions are accounted for, as mentioned above.

The application of the KG formalism to crystalline systems at finite temperatures faces 
severe numerical challenges. First, extended supercells need to be considered (a)~to accurately sample the vibrational degrees of 
freedom in the crystal and (b)~to ensure that all relevant electronic transitions are mapped to vertical excitation in the folded
BZ. Second, dense {\bf k}-grids are needed to converge the Brillouin zone integration in Eq.~(\ref{eq:kg-formula}), even for calculations in an extended supercell featuring a reduced BZ. 
The reason is that in crystalline materials the band structure is typically dispersive,~i.e.,~band energies vary significantly {as function of} {\bf k}. 
As a consequence, the occupation numbers, determined by the Fermi distribution, also show a strong {\bf k}-dependence.
In this regard, also an appropriate representation for numerically evaluating the delta function appearing in Eq.~(\ref{eq:kg-formula}) 
has to be chosen. As discussed below, this can be achieved by a broadening function, whereby it is essential to reach the limit of
vanishing broadening.

In the following, we will discuss these numerical issues by computationally investigating 
mobilities~$\mu$, see Eq.~(\ref{eq:conductivity_defination}). As mentioned in the
introduction, targeting mobilities instead of conductivities allows to disentangle the underlying 
transport mechanisms. Since the conductivity is directly proportional to the amount of free charges,~i.e.,~the 
charge carrier density~$n$, it sensitively depends on doping. Conversely the mobility 
is an intrinsic property of the material that is unaffected by changes in the charge-carrier density in the low 
doping limit. Only for high doping a dependence of mobilities on the charge-carrier density has been observed.
However, even in this case, this is {understood as} being largely caused by the increased defect scattering due to the very high dopant concentration 
and not by changes in the charge-carrier density~\cite{I-te_prm_2022}. For these exact reasons, it is common both in theoretical~\cite{ponce_review,ponce_si_prb} 
and in experimental~\cite{si_carrier_density_mobility_prb_exp, carrier_density_mobility_exp_2} studies of electronic transport 
to report the mobility~$\mu$ and the charge-carrier density separately, instead of just the conductivity. Along these arguments, 
 we henceforth fix the charge-carrier density~$n$ to the desired value in our evaluation of Eq.~(\ref{eq:kg-formula}), as detailed in Sec.~\ref{def_mobility2}. 
Mobilities are then obtained by just dividing the obtained conductivity by~$n$. For DFT calculations, this has the additional advantage that inaccuracies in the 
charge carrier-density, which are 
heavily affected by the notorious band-gap problem of semi-local DFT~\cite{band_gap_problem_dft,band_gap_perdew_pnas}, 
can be circumvented, as also discussed in Sec.~\ref{def_mobility2}.
Note that we will also focus on scalar mobilities,~i.e.,~the average of the diagonal elements:
\begin{equation}
    \mu = \frac{1}{3}\sum_i \mu_{ii}\quad; \quad i \in \{x,y,z\} \;.
\end{equation}

\begin{center}
    \begin{figure}
        \centering
        \includegraphics[width=1\linewidth]{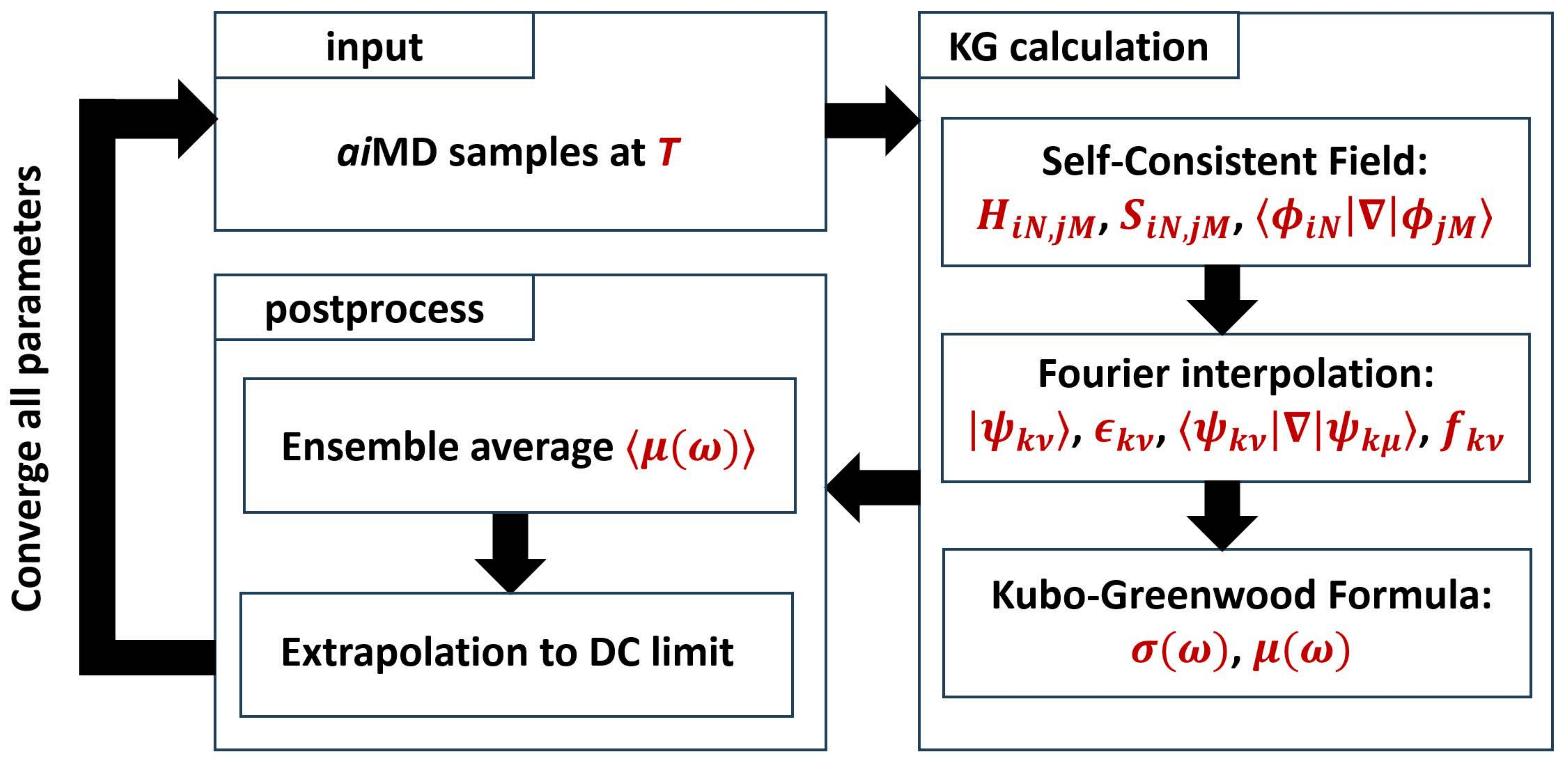}
        \caption{Workflow of the Kubo-Greenwood (KG) calculation. All symbols and steps are explained in details in Section~\ref{sec:implementation_kg}. Convergence here refers to running KG calculations with different parameters until the error in the result is within the desired threshold. } 
        \label{fig:workflow}
    \end{figure}
\end{center}

{A sketch of the {\it ab initio} KG workflow is shown in Fig.~\ref{fig:workflow}. Briefly speaking, the process begins by generating samples via {\it ai}MD simulations at the desired temperature $T$. Subsequently, KG calculation are performed to evaluate the conductivity and mobility of each {\it ai}MD sample. The overall conductivity and mobility at $T$ is then determined by taking the ensemble average over all samples. Details on the numerical convergence schemes employed throughout the process will be thoroughly discussed in the next section.}

\section{Numerical Implementation of the Kubo-Greenwood Formalism}
\label{sec:implementation_kg}
In this work, we implemented the KG formula in the \textit{ab initio} electronic-structure theory package \texttt{FHI-aims}~\cite{fhi-aims}. 
In this all-electron code, the Kohn-Sham states $\ket{\textbf{k}\nu}$ are expanded in terms of (non-orthogonal) 
real-space basis functions $\ket{\phi_{j\textbf{N}}(r)}$
\begin{equation}
    \ket{\textbf{k}\nu} = \frac{1}{\sqrt{L}}\sum_{j\textbf{N}} C_{\nu}^j (\textbf{k}) e^{i\textbf{k}\cdot\textbf{T(N)}}\ket{\phi_{j\textbf{N}}(r)} \;.
\end{equation}
Here, $\ket{\phi_{j\textbf{N}}(r)}$ indicates the basis function~$j$ centered in the unit cell $\textbf{N} = (N_1, N_2, N_3)$, $L$ is the total number 
of periodic images of the unit cell that are accounted for, and $\textbf{T(N)}$ is the translation vector connecting the periodic image in cell~$\textbf{N}$
to the unit cell . Accordingly, the exponential $\exp(i\textbf{k}\cdot\textbf{T(N)})$ ensures that the $\ket{\textbf{k}\nu}$ 
fulfill the Bloch condition. In turn, the momentum matrix elements $i\hbar\braket{\textbf{k}\mu|\nabla|\textbf{k}\nu}$ can be expressed 
in term of the real-space basis functions $\ket{\phi_{j\textbf{N}}(r)}$ as:
\begin{eqnarray}
\braket{\textbf{k}\mu|\nabla|\textbf{k}\nu} = \sum_{ij} &[C_{\mu}^{i}(\textbf{k})]^*C_{\nu}^j(\textbf{k}) \sum_{\textbf{N}} e^{i\textbf{k}\cdot\textbf{T(N)}}\nonumber \\
&\times\braket{\phi_{i\textbf{0}}(r)|\nabla|\phi_{j\textbf{N}}(r)} \label{MomElements}\;.
\end{eqnarray}
Here, $\braket{\phi_{i\textbf{0}}(r)|\nabla|\phi_{j\textbf{N}}(r)}$ denotes the gradient matrix of the NAO basis functions, 
which is evaluated in real-space using the techniques described in Ref.~\cite{stress_aims,dfpt_aims}. Note that $L$ and 
one summation $\sum_{\textbf{M}}$ over unit cells are eliminated in the formula above by exploiting 
the translational symmetry of the real-space matrix elements.

As mentioned above and discussed in details in Sec.~\ref{kConv}, dense \textbf{k}-grids are typically required to reach 
convergence when evaluating the Brillouin zone 
in Eq.~(\ref{eq:kg-formula}). To alleviate the computational cost, we exploit the fact that the electronic density and, in 
turn, the associated real-space Hamiltonian, can typically be converged within a relatively sparse 
set of \textbf{k}-points, especially in extended supercells featuring a finite band gap. 
Formally, this is reflected by the fact that the real-space Hamiltonian and Overlap matrix elements defined 
as $\textbf{H}_{i\textbf{N},j\textbf{M}}=\braket{\phi_{i\textbf{N}}|\textbf{H}|\phi_{j\textbf{M}}}$ and $\textbf{S}_{i\textbf{N},j\textbf{M}}=\braket{\phi_{i\textbf{N}}|\phi_{j\textbf{M}}}$  are localized integrals in the NAO basis. The respective reciprocal-space matrix elements at any $\textbf{k}$ are 
obtained via:
\begin{eqnarray}\label{hk_fourier}
    \textbf{H}_{ij}(\textbf{k}) & = & \sum_{\textbf{M}} 
    e^{i\textbf{k}\cdot\textbf{T(M)}} \textbf{H}_{i\textbf{0},j\textbf{M}} \\
    \textbf{S}_{ij}(\textbf{k}) & = & \sum_{\textbf{M}} 
    e^{i\textbf{k}\cdot\textbf{T(M)}} \textbf{S}_{i\textbf{0},j\textbf{M}} \;.
\label{sk_fourier}
\end{eqnarray}
Due to the locality of the real-space NAO basis, $\textbf{H}_{i\textbf{N},j\textbf{M}}$ and $\textbf{S}_{i\textbf{N},j\textbf{M}}$ 
decay rapidly with atomic distance. Thus, the real-space matrices $\textbf{H}$ and $\textbf{S}$ become sparse, 
since $\textbf{H}_{i\textbf{N},j\textbf{M}}$ and $\textbf{S}_{i\textbf{N},j\textbf{M}}$ vanish at large distances $(\textbf{M}-\textbf{N})$.
In turn, this implies that, once real-space $\textbf{H}$ and $\textbf{S}$ are fully known from the self-consistent field (SCF) calculation 
with a sparse $\textbf{k}$-grid, the respective reciprocal-space $\textbf{H}(\textbf{k}')$ and $\textbf{S}(\textbf{k}')$ at any other 
denser $\textbf{k}'$-grid can be obtained via Eq.~(\ref{hk_fourier}) and Eq.~(\ref{sk_fourier}). With that, it is then possible to 
get the associated Bloch states by solving the generalized eigenvalue problem
\begin{equation}
\textbf{H}(\textbf{k}')\ket{\textbf{k}'\mu}=\epsilon_{\textbf{k}'\mu}\textbf{S}(\textbf{k}')\ket{\textbf{k}'\mu}.
\end{equation}
on denser $\textbf{k}'$-grids. This procedure is often referred to as Fourier interpolation and is commonly used in 
electronic-structure codes for evaluating properties that require dense $\vec{k}$-grids,~e.g.,~band structures and density of states.
In this work, we use Fourier interpolation to evaluate Eq.~(\ref{MomElements}) and hence Eq.~(\ref{eq:kg-formula}) on dense $\vec{k}$-grids, which enables significant memory savings
as well as computational speed-ups and, in turn, enables the practical convergence of the KG equation within reasonable computational cost. 
These merits are discussed in more detail in Appendix.~\ref{appendix_fi}.

\subsection{Charge-Carrier Density Determination}
\label{def_mobility1}
\begin{center}
    \begin{figure}
        \centering
        \includegraphics[width=1\linewidth]{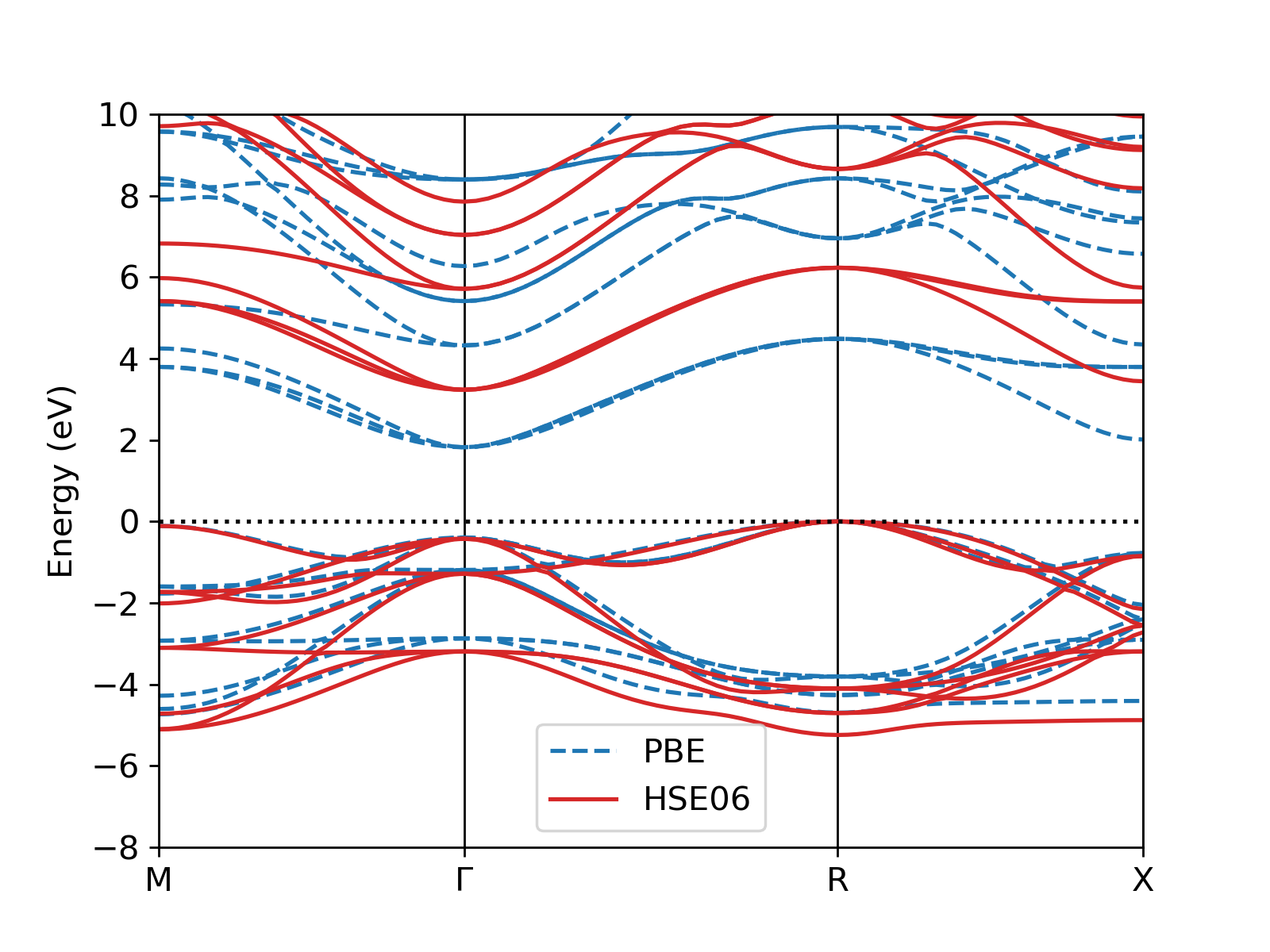}
        \caption{Band structure of the primitive cell of SrTiO$_3$ with different functionals. For both cases, the geometry is that of a PBE-TS calculation. Computational details are introduced in Sec.~\ref{sec: computational_details}.
        }
        \label{fig:band_structure}
    \end{figure}
\end{center}
In the static case,~i.e.,~for a given nuclear configuration, the Fermi level $\rm \epsilon_F$ is determined by enforcing charge neutrality,~i.e.,~by requiring the free electron 
density $ n_e$ in the conduction band~(CB) to match the free hole density $n_h$ in the valence band~(VB):
\begin{equation}
\label{eq:determine_Ef}
{n_e} = \sum^{\rm CB}_{{\bf k},\nu} f({\rm \epsilon_F}, T, \epsilon_{ {\bf k} \nu}) = \sum^{\rm VB}_{ {\bf k},\nu} (1 - f({\rm \epsilon_F}, T, \epsilon_{ {\bf k} \nu}) ) = { n_h} \ ,
\end{equation}
where $f({\rm \epsilon_F}, T, \epsilon)$ is the Fermi distribution function and the spin degree of freedom is omitted.
At finite temperature, $\rm \epsilon_F$ needs to be determined self-consistently, and thus, at the same time also the free carrier density $n_e$ and $n_h$ are known. The Fermi distribution changes exponentially with increasing energy deviation from $\rm  \epsilon-\epsilon_F$.
This indicates that the intrinsic carrier density depends exponentially on the band gap~$E_g$ in zero order approximation: 
\begin{equation}
    {n_e} \propto \exp\left(-\frac{ E_g}{2{\rm k_B} T}\right) \;.
\label{ChargCarrEstim}
\end{equation}
Likewise, the free carrier density can be formally controlled by manipulating the Fermi level $\rm \epsilon_F$ with respect to the condition ${n_e} = {n_h} + n$, 
where $n$ is the desired carrier density.

Due to the well-known band gap problem of semi-local DFT, inaccuracies in the band-gap assessment strongly affect the carrier density and thus the conductivity. For example, 
the PBE band gap for SrTiO$_3$ at $T=0$~K is only~2.11~eV, well smaller than the experimental value of~3.26~eV~\cite{sto_2300k} measured at 10~K. In turn,
this implies that PBE yields intrinsic charge carrier densities and conductivities that are {\it six} orders times too large, as can be estimated with Eq.~(\ref{ChargCarrEstim})
and is explicitly demonstrated later in Sec.~\ref{def_mobility2}.

In principle, accurate conductivity calculations would hence require an appropriate many-body electronic structure theory approach that overcomes the band-gap problem of semi-local DFT,~e.g.,~$GW$.
However, this would be computationally very expensive and as discussed in Sec.~\ref{def_mobility2} below, not really necessary. 
The reason is that, albeit severely underestimating the band gap, semi-local DFT still predicts the curvature/dispersion of the electronic band structure surprisingly well, since also 
the Kohn-Sham wave functions are typically sufficiently accurate~\cite{gw_review,gw_aims}. For the exact same reason, scissor-operator approaches typically 
work well for obtaining free charge carrier densities~\cite{scissor_operator_prb,Zhenkun_2022}.
This is further substantiated in Fig.~\ref{fig:band_structure}, in which the band structure of SrTiO$_3$ computed using PBE and the hybrid HSE06 functional are compared. While HSE06 yields a realistic band gap 
of around 3.52~eV that is comparable to the experimental value of~3.26~eV~\cite{sto_2300k}~at 10~K, PBE severely underestimates the band gap yielding a value of~2.11~eV, as mentioned above. 
Still, the shapes of the individual bands are very similar with both functionals,~e.g.,~the effective masses differ only by $\sim 5\%$,~yielding 0.86~(PBE) and 0.92 $\rm m_e$~(HSE06) 
at $\Gamma$ in M direction. 

\subsection{Convergence with respect to Electronic Degrees of Freedom}
\label{kConv}
As evident from the Kubo-Greenwood formula in Eq.~(\ref{eq:kg-formula}), only transitions between electronic states that differ in occupation number contribute to the conductivity.
In pristine semiconductors with no defect state in the gap, transitions between valence and conduction band hence dominate the optical conductivity for $\hbar\omega$ values that are 
larger than the band gap. When $\hbar\omega$ is smaller than the band gap, however, transitions between valence and conduction band are energetically forbidden by the delta-function
in Eq.~(\ref{eq:kg-formula}). Accordingly, only transitions within the valence band,~i.e.,~between occupied valence states and the free holes in the valence band, or within the conduction 
band,~i.e.,~between unoccupied conduction states and the free electrons in the conduction band, can contribute to the mobility. Discerning these two types of transitions allows to disentangle 
the contributions of holes and electrons to the mobility. Let us note that in the case of (occupied or unoccupied) defect states in the band gap, the exact same arguments hold and the KG formalism
is equally applicable. In this case, however, the relevant energy gap for discerning between hole and electron conductivity is no longer the fundamental band gap between valence and conduction band,
but the gap between the valence band and lowest unoccupied defect state or between the highest occupied defect state and the conduction band. For the sake of simplicity, we will just refer 
to the gap between highest occupied and lowest unoccupied state as transition gap in the following, regardless of defect states being present or not. 

\begin{center}
    \begin{figure}
        \centering
        \includegraphics[width=1\linewidth]{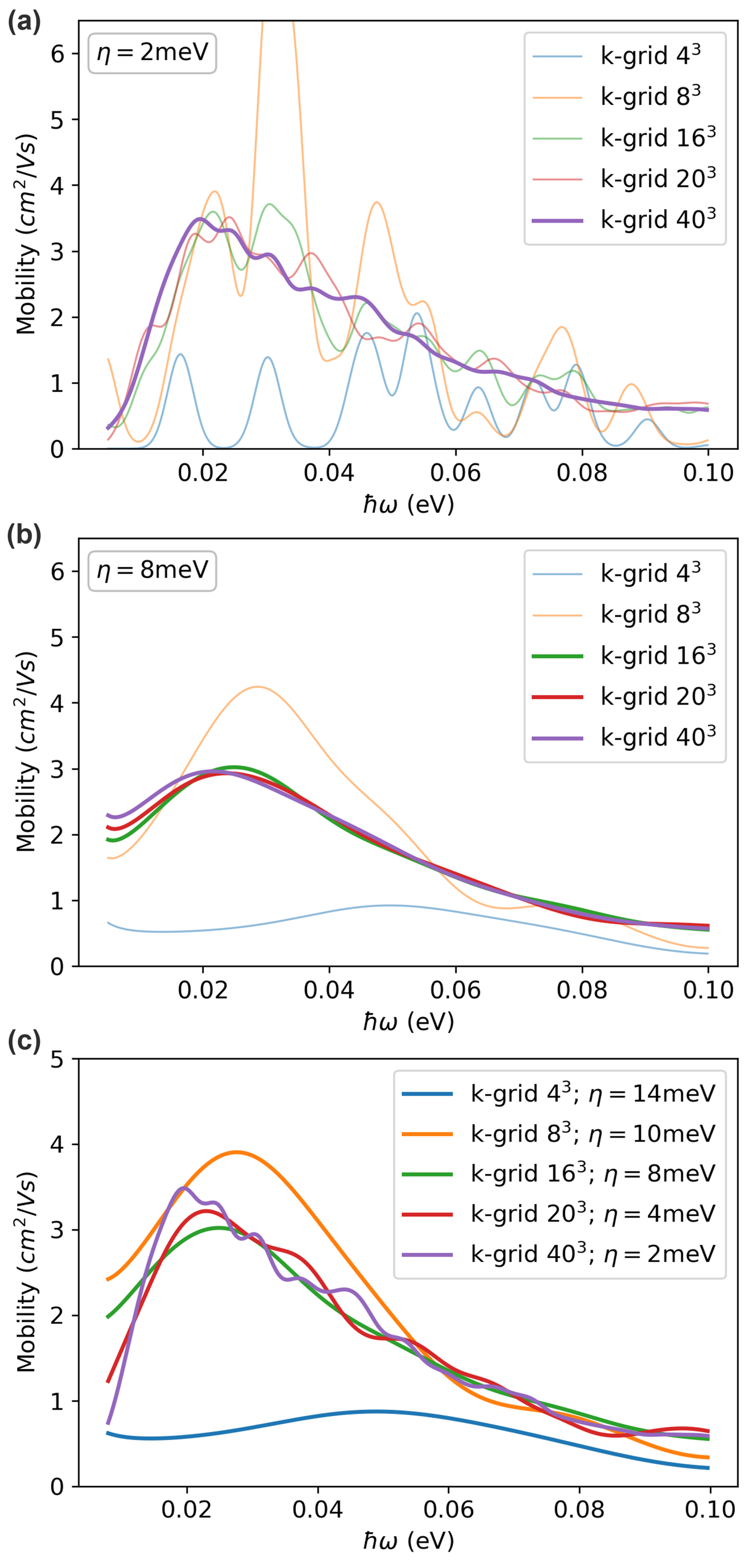}
        \caption{Mobility spectrum calculated from a 40-atom supercell SrTiO$_3$ sample. Convergence of the mobility spectrum with fixed (a) $\eta = 2$meV, (b) $\eta = 8$meV and increasing {\bf k}-grid density. Thicker lines indicate {spectra that can be considered as converged with respect to the {\bf k}-grid density.} In subplot (a), a {\bf k}-grid $40^3$ is needed to converge the spectrum, but in (b) a {\bf k}-grid of $16^3$ is already sufficient. (c) Mobility spectra for different $\eta$ values using individually converged {\bf k}-grid densities.
        }
        \label{fig:k_eta_converge}
    \end{figure}
\end{center}

For the case of $\hbar\omega$ being smaller than the transition gap and the transition gap being larger than $k_BT$,~i.e.,~in case of a non-metallic system,
only those states closest to the Fermi level can contribute, since all deep or high levels far from the Fermi level are all either fully occupied or unoccupied due to the fast decay of the Fermi function.
For semiconductors with dispersive electronic band structures, valence holes and conduction electrons are hence confined to small pockets in \textbf{k}-space near the valence band maximum (VBM), 
the conduction band minimum~(CBM), and, if present, defect states. Accurately sampling these tiny pockets requires very dense \textbf{k}-grids, also because the actual location of these pockets is not
known a priori, since the band structure and these pockets depend on the atomic positions.
The use of Fourier interpolation gives access to dense \textbf{k}-grids in a numerically efficient and reliable fashion, so that one can reach convergence, as shown in Fig.~\ref{fig:k_eta_converge}(a) 
for one selected 40 atoms SrTiO$_3$ sample from aiMD trajectory at 500K. When the {\bf k}-grid is too sparse, e.g.,~for a $\vec{k}$-grid of $4^3$, there are large oscillations in the spectrum, since transitions within the
pockets are not accurately sampled. For increasingly  dense \textbf{k}-grids, the spectrum becomes smooth and convergence is observed, here for grids around $40^3$.

Note that the converge behavior with respect to the $\vec{k}$-grid also sensitively depends on the numerical representation of the delta-function in the evaluation of Eq.~(\ref{eq:kg-formula}).
In this work, a Gaussian broadening 
\begin{equation}
    G_{\eta}(\omega) = \frac{1}{\sqrt{\pi}\eta} \exp{\left(-\frac{(\epsilon_{\textbf{k}\mu}-\epsilon_{\textbf{k}\nu}-\hbar\omega)^2}{\eta^2}\right)}
\end{equation}
is used,  but our findings are not limited this particular broadening function. For the example in Fig.~\ref{fig:k_eta_converge}(a),
a converged broadening width of $\eta=2$~meV was used. As discussed in Sec.~\ref{sec:theory_kg}, convergence is reached for $\eta$ in 
the zero broadening limit. To approach this limit, the \textbf{k}-grid has to be chosen the denser, the smaller the broadening parameter $\eta$ is, since sharper broadening implies less BZ averaging. 
This can be seen by comparing Fig.~\ref{fig:k_eta_converge}(a) and (b), in which the exact same $\vec{k}$-grids, but different values for $\eta$ = 2 and 8~meV are used. In the later case, $\vec{k}$-grid $16^3$ is already converged.

Eventually, we discuss the convergence behaviour with respect to $\eta$, whereby the $\vec{k}$-grid was converged individually for each value of $\eta$. For broadenings larger than $\eta>10$~meV, {\bf k}-grid 
convergence is already achieved with sparse grids. However, the shape and values of these spectra are unreliable,~i.e.,~they are largely determined by the broadening function 
and not by the underlying physics. As shown in Fig.~\ref{fig:k_eta_converge}(c), 
the spectra convergence for  lower values of $\eta$ and the results do no longer depend on the chosen value of $\eta$. However, considerable denser \textbf{k}-grids are needed
to reach this limit. In this case of a 40 atom SrTiO$_3$ cell, a broadening of $\eta=4$~meV and a \textbf{k}-grid of $20^3$ can be considered as converged. 

Obviously, the value of $\eta$ and the \textbf{k}-grid density 
needed to achieve convergence is system-dependent. In general, sparser \textbf{k}-grid are sufficient if the dispersion becomes flatter, since then free holes and conduction electrons are less
localized in pockets of the BZ. This also explains why convergence is more easily achieved in disordered systems, as mentioned in the introduction.

\subsection{Convergence with respect to Nuclear Degrees of Freedom}
\label{sec: thermodynamic_average}

\subsubsection{Thermodynamic average over samples}

\begin{center}
    \begin{figure}
        \centering
        \includegraphics[width=1\linewidth]{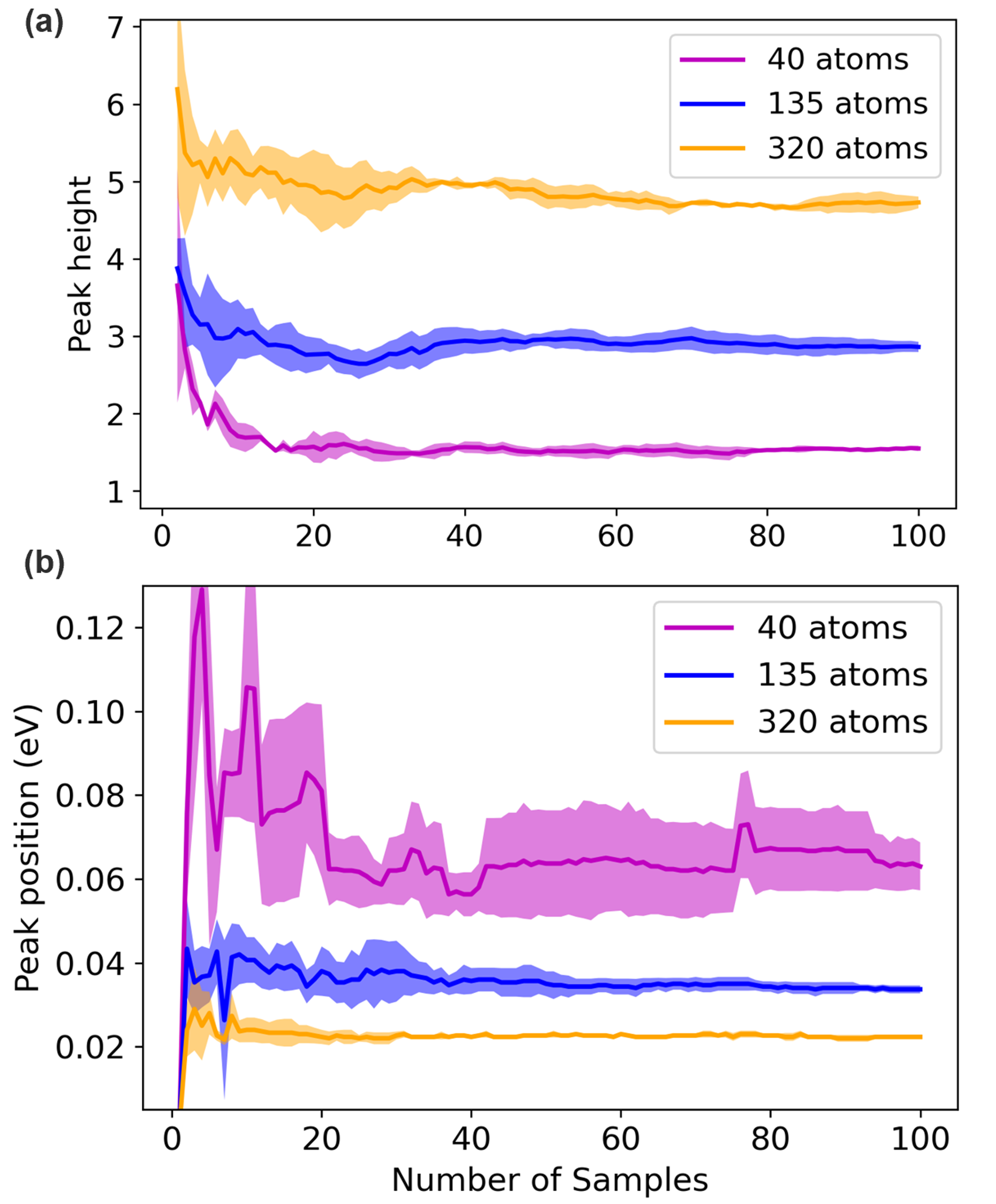}
        \caption{Evolution of the averaged (a) peak height and (b) peak position over 3 different set of samples as function of the considered number of samples (x-axis) in each set. Samples were randomly selected from one 4ps {\it ai}MD trajectory. The dashed region indicates the standard deviation over the 3 set of samples. 
        }
        \label{fig:ensemble_peak_pos}
    \end{figure}
\end{center}

\begin{center}
    \begin{figure}
        \centering
        \includegraphics[width=1\linewidth]{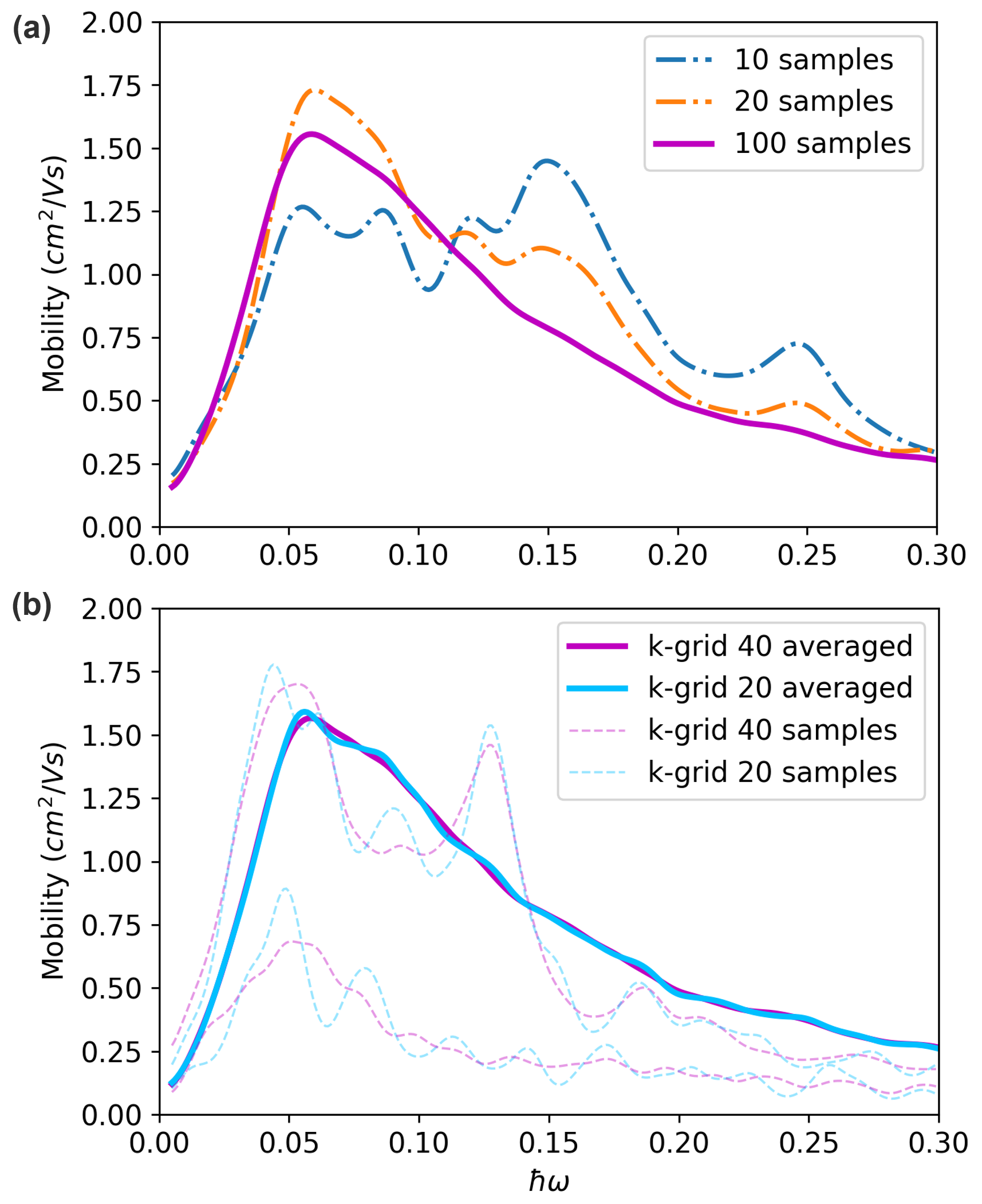}
        \caption{(a) Mobility spectra for the 40-atom supercell SrTiO$_3$ obtained by KG-averaging samples from an {\it ai}MD trajectory. Subplot~(a) shows spectra averaged over different number of
samples, while~(b) compares spectra averaged over 100 samples and spectra of one specific sample computed with different k-grid densities. 
        }
        \label{fig:ensemble_samples}
    \end{figure}
\end{center}

As discussed in the context of Fig.~\ref{fig:workflow}, computing the mobility requires an average over multiple mobility spectra associated to different geometric configurations, so to cover the relevant phase space accessible at a specific temperature. To monitor the convergence of the spectrum with respect to phase space sampling, Fig.~\ref{fig:ensemble_peak_pos}(a) shows  
the peak height of the averaged spectrum as function of the number of samples,~i.e.,~of the number of different geometric configurations used for averaging, for different
supercell sizes.
In practice, samples are randomly selected from an {\it ai}MD trajectory, cf.~Sec.~\ref{sec: computational_details} for details, and grouped in three different sets with
the same sample size, which are used to determine averages and standard deviation of the average across sets.
By only considering the peak height, one might conclude that the spectra converge very quickly,~e.g.,~the 40 atom supercell 
seems already converged with less than 20 samples. However, further analysis reveals that this is not the case. As shown in Fig.~\ref{fig:ensemble_samples}(a), 
the overall shape of the spectrum changes quite dramatically when further increasing the number of samples to a 100, even if the peak height
does not. To take this into account, we also plot the evolution of the peak position in Fig.~\ref{fig:ensemble_peak_pos}(b), which better reflects the
convergence of the spectrum's shape. This also reveals that the peak position converges more rapidly for larger supercell size, since more phase space is 
sampled in larger supercells. Accordingly, converged spectra for the 40, 135 and 320-atom supercells are achieved with at least 100, 50 and 30 samples, respectively.

{Not too surprisingly, we find that} the thermodynamic averaging effectively smoothens the spectrum, as showcased in Fig.~\ref{fig:ensemble_samples}(a). In turn, this implies that --compared to the spectrum of a single sample-- less dense {\bf k}-point grids are needed to converge the thermodynamically averaged spectrum. This is also seen in Fig.~\ref{fig:ensemble_samples}(b), which shows that the averaged spectrum for \textbf{k}-grids of $20^3$ and $40^3$ are essentially equal. In comparison, the respective spectra obtained with \textbf{k}-grids of $20^3$ and $40^3$ for a single sample do not agree and show differing oscillation patterns, as discussed in the previous section and shown in Fig.~\ref{fig:k_eta_converge} and dashed lines of Fig.~\ref{fig:ensemble_samples}(b).

\subsubsection{Thermodynamic Averages of Charge-Carrier Densities}
\label{def_mobility2}
\begin{center}
    \begin{figure}
        \centering
        \includegraphics[width=1\linewidth]{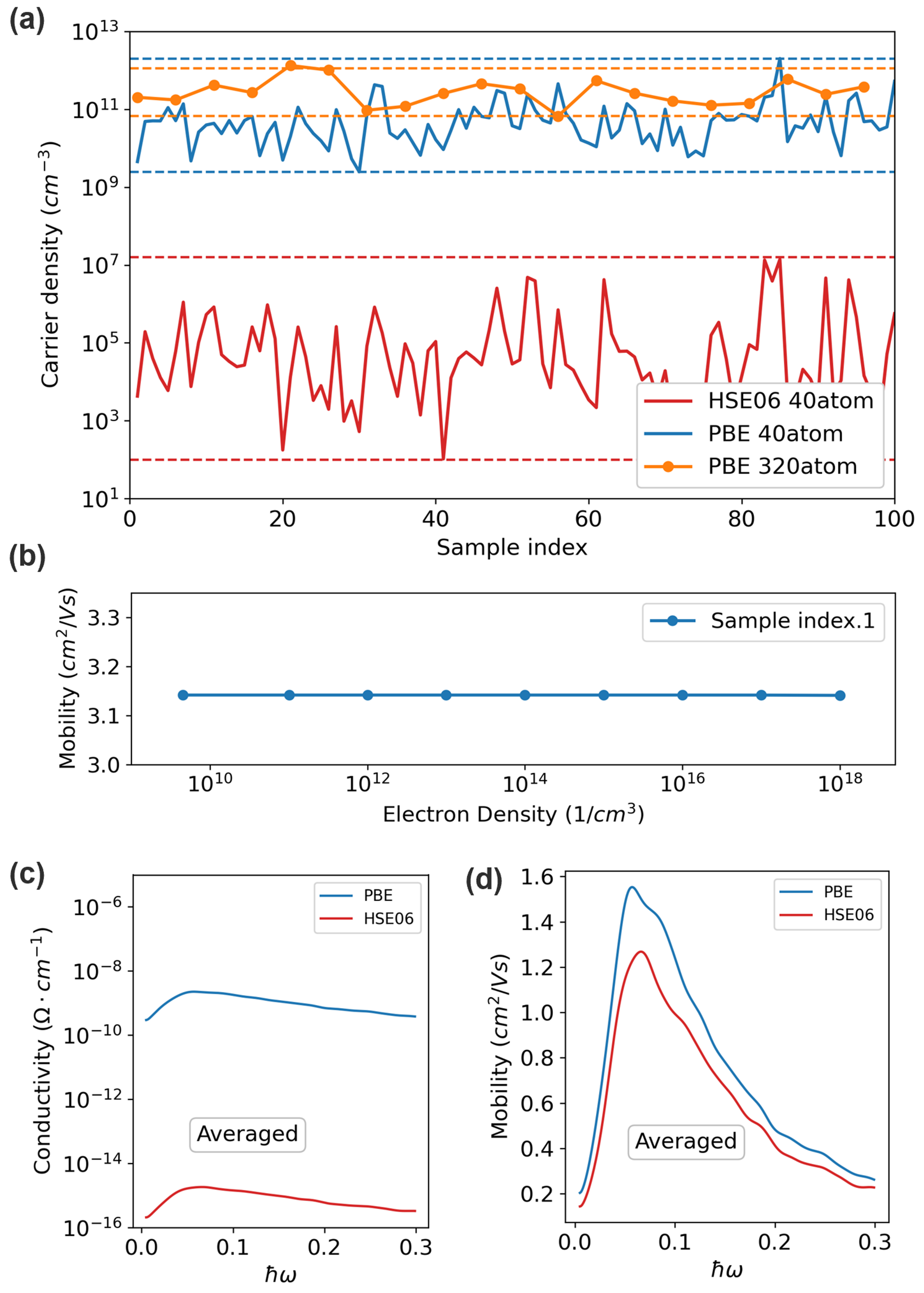}
        \caption{(a) Fluctuation of carrier density in each sample with different functional and supercell sizes. (b) Height of the first peak in the mobility spectrum for different fixed charge carrier densities~(one sample, 40-atom supercell) (c-d) Comparison of averaged conductivity/mobility spectrum for the 40-atom supercell and different functionals over 100 samples. 
        }
        \label{fig:set_carrier_density}
    \end{figure}
\end{center}
As discussed in Sec.~\ref{def_mobility1}, the charge carrier density sensitively depends upon the band gap. 
Fig.~\ref{fig:set_carrier_density}(a) shows the free electron densities $n_e$ obtained for individual samples with PBE and HSE06,
resulting in average intrinsic $\braket{n_e}$ that is nearly 6 orders of magnitude too large with PBE compared to HSE06 due to the 
underestimation of the band gap by PBE. In turn, this results in conductivities that are also nearly 6 orders of magnitude
larger with PBE than with HSE06, see Fig.~\ref{fig:set_carrier_density}(c). Targeting mobilities,~i.e.,~separating out the contribution stemming
from the charge carrier density, allows to obtain much more reasonable results already at the PBE level. However, one further subtle aspect 
arises during the thermodynamic averaging of conductivities and charge carrier densities in the determination of mobilities. 
In experiments,  the mobility is obtained by measuring conductivity and charge-carrier density separately,~i.e.,~via $\braket{\mu} = \frac{\braket{\sigma}}{\braket{qn_c}}$. 
Computationally, this route is in principle also viable, since ${\braket{qn_c}}$ can be obtained by averaging over the individual samples, which all feature
 a different atomic configuration and, in turn, a different electronic structure, intrinsic Fermi energy, and carrier density. In that case,
one obtains
\begin{eqnarray}
  \mu =  \frac{\braket{\sigma}}{\braket{qn}} &=& \frac{ \sum_i \sigma_i }{ q \sum_i n_i }  =  \sum_i \frac{n_i}{\braket{n}} \mu_i\;,
\label{eq_avg}
\end{eqnarray}
in which the sum over~$i$ runs over all samples. While correct in the thermodynamic limit, Eq.~(\ref{eq_avg}) leads to numerical problems for microscopic systems 
that exhibit artificially large charge carrier densities fluctuations. As shown in Fig.~\ref{fig:set_carrier_density}(a), the charge carrier densities fluctuate 
massively across samples in finite systems, since even just small changes in the band gap lead to exponential changes of the carrier density. For the 40 atom supercell, 
for instance, the PBE values vary over three, the HSE06 values over five orders of magnitude, showcasing that these fluctuations are even more pronounced 
in case of higher band gaps and lower charge carrier densities. In turn, this leads to the undesirable side-effect that only very few samples with high charge carrier 
density determine the average in Eq.~(\ref{eq_avg}) and that noise cancellation via averaging is largely suppressed so that the spectra converge very slowly with the
number of samples. 

The above mentioned issues are particularly problematic when comparing different supercell sizes. For larger, but still microscopic cells,~e.g.,~the 320~atom cell also 
shown in Fig.~\ref{fig:set_carrier_density}(a), the flucations are already massively attenuated and ``only'' cover one order of magnitude in the case of PBE. For larger systems,
when approaching the themodynamic limit, these fluctuation would be even more suppressed. To correctly reflect the thermodynamic limit already in small cell sizes, it is thus 
advantageous to not use Eq.~(\ref{eq_avg}), but to eliminate these fluctuations by fixing the charge-carrier density across samples. In practice, this is achieved by shifting 
the intrinsic Fermi level for each individual sample to achieve the desired carrier density during the Kubo-Greenwood calculations. 
To this end, one can for instance fix the charge-carrier density to be close to the intrinsic limit or to the experimentally measured value. By these means, it is 
ensured that the charge-carrier density of interest is addressed and that the averaging is numerically reasonably efficient. However, this approach efficitively eliminates all
fluctuations, even those minute ones that would still be present in the thermodynamic limit. We find that this approximation has negligible influence, since the mobilities of 
each individual samples are largely independent of the carrier density in a wide range, a shown in Fig.~\ref{fig:set_carrier_density}(b). 
Eventually, let us note that the developed strategy holds in the intrinsic and low-doping limit. For high dopant concentrations, an explicit simulation of the dopants within 
the Kubo-Greenwood approach is also straightforward. 

In practice, the proposed approach minimizes the error stemming from the usage of semi-local DFT and of the thereby underestimated band-gap, as alluded before. 
As shown in Fig.~\ref{fig:set_carrier_density}(c), the conductivity with PBE is $\sim 10^6$ times larger than the one of HSE06, if one does not fix the charge carrier density.
In comparison, our approach with fixed charge carrier densities shown in Fig.~\ref{fig:set_carrier_density}(d) yields similar mobilities for PBE and HSE06. For instance, the peak height 
only differs by $\sim 13\%$, in line of what can be expected due to different band curvatures obtained with different functionals~\cite{ponce_si_prb,ponce_review,yuanyue_2d_mobility}

\begin{center}
    \begin{figure}
        \centering
        \includegraphics[width=1\linewidth]{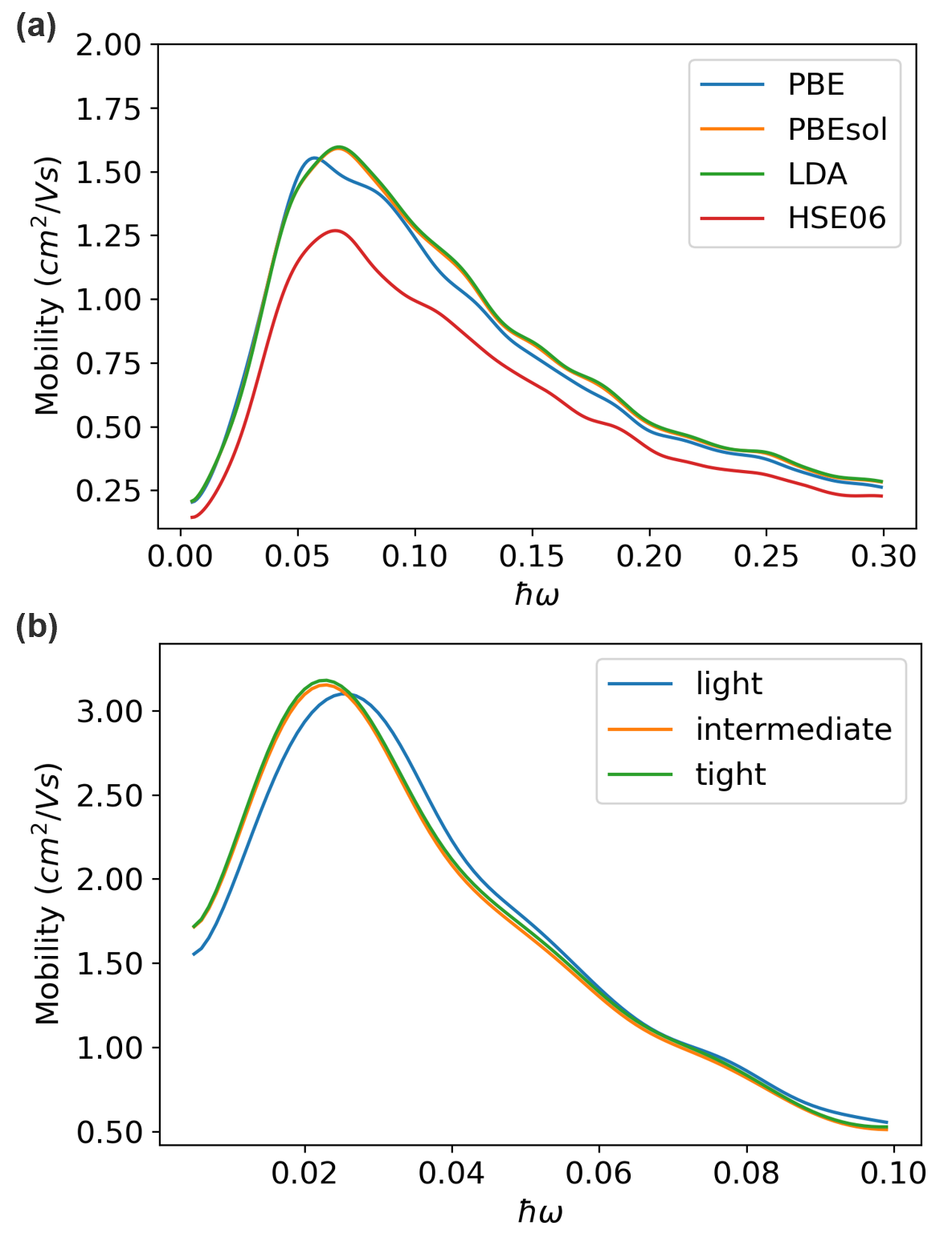}
        \caption{(a) Averaged mobility spectrum for the 40-atom SrTiO$_3$ supercell~(100 samples, {\it ai}MD simulations performed with light settings and the PBE functional) with different xc functionals. (b) Mobility spectrum of one sample calculated with PBE functional and different numerical settings.}
        \label{fig:xc_basis}
    \end{figure}
\end{center}

Besides PBE and HSE06, we also tested other xc functionals since they give different effective single-particle states, different band curvatures, and hence different mobilities. As shown in the Fig.~\ref{fig:xc_basis}(a), the local-density approximation (LDA) with the PW92 functional~\cite{pw92} and the generalized gradient approximation (GGA) with the PBE~\cite{pbe_functional} and PBEsol~\cite{pbesol} functionals give very similar results with fixed carrier density. Besides the xc functionals, also all other numerical settings need to be converged, including the basis sets. Default numerical settings 
in {\tt FHI-aims} are divided into {\it light}, {\it intermediate}, {\it tight}, and {\it really\_tight}. ``Tighter'' settings imply a higher number of NAOs, basis functions with longer-range tails,
denser integration grids, and more~\cite{fhi-aims,Carbogno.2022}. As shown in Fig.~\ref{fig:xc_basis}(b), {\it light}, {\it intermediate}, and {\it tight} settings give very similar results for the 40-atom SrTiO$_3$ cell, so that the computationally most affordable {\it light} settings are used for the remainder of the paper.
 
\subsection{Finite-size effect and Extrapolation}
\label{sec:extrapolation}
\begin{center}
    \begin{figure}
        \centering
        \includegraphics[width=1\linewidth]{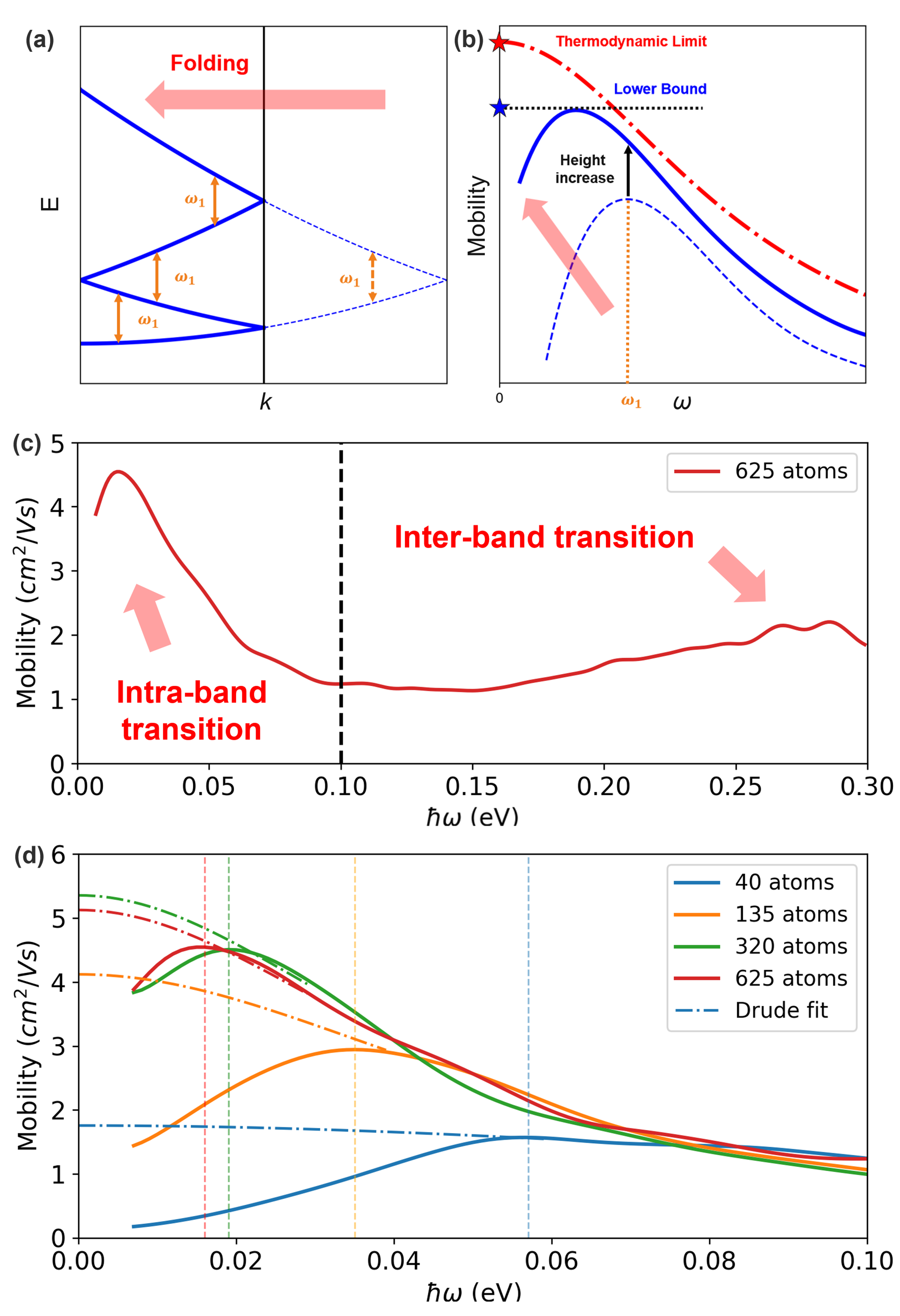}
        \caption{(a)Schematic band structure to show the effect of band folding with increasing supercell size. Dashed lines indicate the bands of a small unit cell, where a direct transition with energy difference $\hbar\omega_1$ can only happen at a single k-point. When the supercell size is doubled, transitions become possible at three k-points. (b) Schematic mobility spectrum to illustrate the increase of peak height and shift of peak position with increasing supercell size. (c) The mobility spectrum of a 625-atom SrTiO$_3$ at 500K with inter-band peaks at higher frequencies. The black dashed line indicates the upper-bound of the frequency for a reasonable Drude fit. (d) The mobility spectrum of SrTiO$_3$ at 500K obtained for different supercell sizes. The thin dashed vertical line indicates the peak position. It shows that the peak position approaches zero as the supercell size increase, so that the Drude fit gradually converges. }
        \label{fig:supercell_converge}
    \end{figure}
\end{center}

Eventually, we analyse the convergence with respect to the supercell size, which largely determines the behaviour of the optical conductivity in the low frequency limit. 
As mentioned in Sec.~\ref{sec:theory_kg}, the use of extended supercells is required to correctly sample long wavelength vibrational modes, which in turn allows to sample
indirect transitions,~i.e.,~electronic transitions between states at different $\vec{k}$-points. This happens via Brillouin zone folding, which maps transitions that are
originally indirect in the first BZ associated to the primitive cell to direct transitions in the reduced BZ associated with a supercell. In finite supercells, only a fraction of indirect transitions 
is mapped and accounted for. This is evident from Eq.~(\ref{eq:kg-formula}), in which energy conservation requires that the energy difference for a transition between two states has to equal~$\hbar\omega$. As sketched in Fig.~\ref{fig:supercell_converge}(a), there are only few states with small energy differences $\hbar\omega$ in the band structure of a small supercell. 
Thus, there are no direct transitions for small values of $\hbar\omega$. Accordingly, the respective spectrum features a peak and then quickly decays to zero for small values of $\omega$. 
For increasing supercell sizes, the BZ is folded, as shown in Fig.~\ref{fig:supercell_converge}(a). Hence, more vertical transitions with smaller energy difference become 
possible and are naturally included in Eq.~(\ref{eq:kg-formula}). As a result, the peak in the mobility spectrum becomes higher and its position shifts closer towards $\omega = 0$, as already
observed in literature~\cite{plasma_pre_2022}. In this context, it is important to mention that Fröhlich coupling,~i.e.,~electron-phonon coupling in the long wavelength 
limit~\cite{giustino_rmp,frohlich_giustino}, is obviously hard to capture within finite-size supercells and is thus neglected in this work.

The above described trends can also be rationalized in terms of ``unfolded'' electronic and vibrational reciprocal-space states,~i.e.,~in the primitive unit cell and BZ. Small supercells only feature few vibrational 
modes characterized by a coarse reciprocal-space resolution~$\Delta\vec{q}$ of the vibrational degrees of freedom in the primitive BZ. In turn, only indirect transitions that fulfill crystal momentum conservation~$\vec{k}+\vec{k}'=\Delta\vec{q}$ 
are sampled. For increasing supercell size, more long-wavelength vibrations are included and the resolution of the vibrations in reciprocal space improves,~i.e.,~$\Delta\vec{q}$ decreases. Accordingly, more indirect transitions
between $\vec{k}$-points that are close to each other are accounted for. Since such states at ``neighbouring'' $\vec{k}$-points typically feature smaller energy differences, they play an exceedingly important role for the
low-frequency limit~$\hbar\omega\to 0$. 
In an infinitely large supercell, all possible vibrational modes would be included and the mobility would smoothly approach its peak at the DC limit, as sketched in Fig.~\ref{fig:supercell_converge}(b).
In this case, the artificial, numerical divergence at $\omega=0$, which originates from the ${1}/{\omega}$ factor in Eq.~(\ref{eq:kg-formula}), is also lifted. 
For a single band, only intra-band transitions are possible. As we enlarge the supercell size, the BZ folding effect maps these intra-band (indirect) transitions in the primitive cell (PC) to inter-band (direct) transitions in the supercell (SC) as shown in Fig.~\ref{fig:supercell_converge}(a). However, in real multi-band systems, increasing the supercell size can also enable some previously forbidden inter-band (indirect) transitions in the PC to occur, leading to the emergence of new peaks as the frequency increases as shown in Fig.~\ref{fig:supercell_converge}(c). These inter-band transitions w.r.t the PC bands are important for calculating optical properties~\cite{claudia_optical_CPC}, but do not contribute to the mobility in the DC limit.

In practical calculations, however, 
the above mentioned ``infinite supercell'' limit can hardly be reached, as shown in Fig.~\ref{fig:supercell_converge}(d). It is thus important to employ strategies to extrapolate to the DC limit, several of which have been proposed in literature~\cite{vasp_kg_2002,vasp_al_test_2013,abtew_dc_limit,modified_drude_fit_2021,qe_kg_cpc}.

In this work, we extrapolate to the DC limit by fitting the frequency-dependent mobility with a Drude function~\cite{vasp_kg_2002,qe_kg_cpc,drude_fit_prb_2005}. 
Formally, this can be justified  directly from Eq.(\ref{eq:kg-formula}). With a finite broadening and when neglecting inter-band transitions, the conductivity spectrum becomes a Drude function 
\begin{equation} \label{eq:drude_function}
    \mu(\omega\to0) \approx \frac{\mu_0}{(\omega\tau)^2 + 1}
\end{equation}
in the low-frequency limit,  whereby $\mu_0$ corresponds to the DC limit of the mobility and $\tau$ is an effective lifetime of the charge carriers, as shown in Ref.~\cite{qe_kg_cpc}.
Formally, the Drude model was first developed to describe metallic conduction; in practice, the low-frequency optical conductivity of many semiconductors exhibits a Drude-like behaviour,
as long as the additional charge carriers are indeed metal-like conduction-band electrons or valence-band holes, but not localized polarons~\cite{drude_semiconductor_exp,drude_semiconductor_review,sto_mass_enhance_prl}.

The advantage of this approach is that minimal assumptions regarding the behaviour of the mobility spectrum are required. In turn, however, it requires well-converged
mobility spectra with respect to $\vec{k}$-grid, broadening~$\eta$, thermodynamic averages, and supercell size. Achieving this convergence is possible in our implementation 
due to the advancements described in the previous sections.
 
In practice, it implies that we fit the (already thermodynamically averaged) mobility spectrum with Eq.~(\ref{eq:drude_function}) in the low frequency region using an appropriate frequency window. 
{We note that} (a)~the very-low-frequency limit is not trustworthy, as discussed above, and (b)~the high-frequency regions features peaks stemming from inter-band transitions not captured
by a Drude model, as shown in Fig.~\ref{fig:supercell_converge}(c). The inter-band transitions manifest as additional peaks in the mobility spectrum as frequency increase. These peaks enlarge as temperature and supercell size increase since more inter-band transitions occur. Accordingly, we choose a window between 
the first peak and $\sim 40$ meV afterwards. For instance, the fitting in Fig.~\ref{fig:supercell_converge}(d) is performed between 0.015~eV and 0.06~eV for the 625 atom supercell. 
The actual outcome of the
fitting is rather insensitive to the actual specific of the window, as long as it follows the above criteria. For instance, enlarging the window towards higher frequencies of up to~0.1~eV or narrowing the window by only starting the fit at 
0.025~eV only changes the DC mobility by 3\%.

This procedure enables systematic convergence with respect to supercell size, since the fitting window is adapted to the employed supercell size through the location of the first peak in the spectrum. As the example in
Fig.~\ref{fig:supercell_converge}(d) shows, the obtained Drude fits converge with supercell size. In turn, the DC mobilities and lifetimes only vary by 4.3\% and 1.9\% for the largest cell sizes considered in this example. 
Clearly, the employed Drude fitting is an approximation that is desirable to be overcome by more explicit models based on first-principles modeling.
However, it at least allows for systematic convergence, as the numbers above show, and enables an error estimate, given that the peak height 
observed in the largest possible supercell can serve as a lower bound for the mobility in the bulk limit.

The supercell size convergence is highly system dependent. For some materials, an extremely large supercell may be necessary for convergence while for others, a much smaller one might suffice. Take for example, an extreme case of a system with completely flat bands. In such scenarios, increasing supercell size will not map indirect transitions into direct ones, and all k-points are equivalent. Therefore, only one single k-point and a primitive cell is sufficient to converge the spectrum. From this, a simple rule of thumb of supercell size convergence could be: for materials with flatter bands, lower mobilities or those studied at very high temperatures, a smaller supercell is generally sufficient for convergence.

\section{Application}
\label{sec: application_kg}

{In this section, we use the the Kubo-Greenwood method to calculate temperature-dependent electron mobility of the highly anharmonic oxide perovskites SrTiO$_3$ and BaTiO$_3$ up to high temperature and compare with other theoretical and experimental values. Then we further analyze our results in terms of temperature-dependent spectral functions.}

\subsection{Computational Details}
\label{sec: computational_details}
All our calculations were performed with the all-electron, numerical atom-centered orbital basis {\it ab initio} simulation code {\tt FHI-aims}~\cite{fhi-aims}. 
To describe exchange and correlation, we employ the Perdew-Burke-Ernzerhof (PBE) functional~\cite{pbe_functional} when using the generalized gradient approximation (GGA) approximation 
and the Heyd-Scuseria-Enzerhof~(HSE06) functional~\cite{hse_03,hse_06} for hybrid DFT calculations. The efficient hybrid DFT implementation in \texttt{FHI-aims} is described in Ref.~\cite{aims_hybrid_functional_2024, aims_hybrid_functional_2015}. The Tkatchenko-Scheffler~(TS) method~\cite{ts_vdw} is used to take into account 
van der Waals interactions.

We first optimized the lattice constants at 0K with a $5\times 5\times 5$ $\textbf{k}$-grid. Then we perform Born-Oppenheimer {\it ab initio} molecular dynamics ({\it ai}MD) at different temperatures and supercell sizes in the canonical ensemble~(NVT) using a time step of 1~fs for a total of 4,000 steps~(4~ps). After the first 2~ps, the lattice parameters are adjusted such that thermal pressure is minimal to take into account the lattice thermal expansion~(LTE)~\cite{florian_gk_implementation}. The last 2~ps of the aiMD trajectory were then used to evaluate the KG formula. For the {\it ai}MD trajectory of the 625-atom  SrTiO$_3$ supercell, the trajectory computed in Ref.~\cite{marios_bzu} using the same code and computational settings was used. All other {\it ai}MD simulation and post-processing (calculate anharmonicity metric, etc.) are performed via {\tt FHI-vibes}~\cite{fhi-vibes}, whereby {\tt Phonopy}~\cite{phonopy-JPCM} is used as calculator for the harmonic force constants.

For the KG calculations, we use the same k-grid density ($40\times 40\times 40$ $\textbf{k}$-points in the first Brillouin zone of the 5-atom primitive structure) 
for all supercell sizes (i.e. $20\times 20\times 20$ $\textbf{k}$-grid for a 40-atom supercell, etc.). A carrier density of $n_e = 1\times10^{18} (1/cm^3)$ 
is used as discussed in Sec.~\ref{def_mobility2}. Frequency-dependent mobilities are calculated for $\hbar\omega$ values up to 0.3~eV with a resolution of 1~meV 
using a Gaussian broadening function with broadening parameter $\eta\in [1, 10]$ meV~for the delta function in the KG formula. 
The final result for different supercells use the smallest possible $\eta$, such that the spectrum is converged: For the 40, 135, 320 and 625 atom supercells, $\eta=4, 3, 2$ and 2 meV are used as converged values, respectively. Considering empty bands up to 1~eV above the Fermi level is enough to converge the conductivity in each sample due to the fast decay of the Fermi distribution function, as discussed at the beginning of Section.~\ref{kConv}. The number of samples to converge the mobility is ensured for all our calculations. For the 625 atom supercell, 30 samples are for instance enough to converge the result. 

The spectral functions are calculated by unfolding supercell band structures taken from {\it ai}MD samples into the primitive cell Brillouin zone. The averaged unfolding weights over samples are defined as the non-perturbative spectral function taken into account all anharmonic vibration couplings~\cite{marios_bzu}. This non-perturbative spectral function from unfolding is equivalent to a Feynman expansion to all orders in the perturbation~\cite{nery_spectralfunction_feynmandiagram}. The asymmetricity metric $\gamma$ of the spectral function is defined by the integration:
\begin{equation} \label{eq:asymmetric_metric}
    \gamma = \frac{\int f(\omega) (\omega_{peak} - \omega) d\omega}{\int f(\omega) d\omega}
\end{equation}
where $\omega_{peak}$ is the peak position of the spectral function and $f(\omega)$ is the spectral weight at $\omega$. For a perfectly symmetric function centered at $\omega_{peak}$, $\gamma$ will be zero. For spectral functions with multiple peaks with similar height, $\omega_{peak}$ is chosen as the mean position of the highest two peaks.

Impurity scattering of dopants is estimated via the semi-empirical Brooks-Herring model~\cite{brooks1951,brooks_herring_model_1977}:
\begin{equation} \label{eq:defect_formula}
    \mu_{i} = \frac{2^{7/2}\varepsilon^2(k_BT)^{3/2}}{\pi^{3/2}e^3\sqrt{m^*_d}n_{i}G(b) } ,
\end{equation}
with $G(b)=\ln(b+1)-b/(b+1)$, $b=24\pi m^*_d\varepsilon(k_BT)^2 / e^2h^2n'$ and $n'=n(2 - n/n_{i})$. Here, $\varepsilon$ is the dielectric constant, $m^*_d$ is the density-of-state effective mass, $n$ is the free carrier density and $n_i$ is the dopant density, in our calculation we assume that $n=n_i$. For SrTiO$_3$, we use the calculated $\varepsilon = 5.97$ and the experimental electron density $n=1.4\times 10^{18} cm^{-3}$~\cite{sto_experiment}, $m^*_d = 1.8 m_e$~\cite{m_d_sto_1.8}. For BaTiO$_3$, we use the calculated $\varepsilon = 6.40$ and the experimental electron density $n=8.5\times 10^{18} cm^{-3}$~\cite{bto_mobility_400k_exp}, $m^*_d = 6.5 m_e$~\cite{bto_mobility_400k_exp}. The total mobility is then estimated by the Matthiessen rule: $\mu_{tot}^{-1} = \mu_{KG}^{-1} + \mu_i^{-1}$.

\subsection{Strong Anharmonic Effects in Perovskites}

Perovskite materials, characterized by the formula $ABX_3$, have consistently garnered significant interest across various research fields~\cite{perovskite_solar_cell_review,oxide_perovskite_review,oxide_perovskite_review_2}. These materials typically exhibit low-symmetry tetragonal or orthorhombic structures at lower temperatures. As the temperature increases, their structural symmetry increases, 
and phase transitions to high-symmetry cubic structures with a single formula unit per cell are common. These cubic and paraelectric phases of perovskites, featuring untilted octahedra,
were confirmed by X-ray diffraction~\cite{MAPbI3_xrd_cubic}.
However, multiple experiments find local centrosymmetry broken and local disorder in many cubic phase perovskites ~\cite{MaPbI3_distortion_exp,bto_distort_shg,pdf_in_bto,pyroelectric_in_bto}. Similarly, on the theory side, cubic perovskites often have imaginary phonon frequencies, which indicate that these structures do not corresponds to a local potential-energy minimum~\cite{marios_bzu,marios_perovskites_npj}. In turn, a structure optimization of supercells result in structures with lower symmetry~\cite{perovskites_distortion_xgZhao_MaterialsToday}. With all this evidence, researchers find that in reality the cubic paraelectric phase of halide and oxide perovskites is a thermodynamic average of locally disordered octahedral motifs, associated with a potential-energy surface~(PES) featuring multiple wells~\cite{polymorphous_of_IOHP_prb,perovskites_distortion_prb,bto_distortion_xgZhao,marios_perovskites_npj}. Similar effects are common to many oxides, and have, for example been observed in ZrO$_2$~\cite{Fabris.2001,Carbogno.2014}. This corrugated PES results in strong anharmonic effects with respect to the nuclear vibrations, which will strongly affect thermal and charge transport~\cite{low_kappa_perovskites,anharmonic_mobility_perovskites}. Here we choose the two typical oxide perovskites SrTiO$_3$ and BaTiO$_3$ as examples to show the influence of the anharmonic dynamics on the electron mobility.

SrTiO$_3$ (STO) is one of the most widely studied materials in both the oxides and the perovskites families. At low temperatures, it shows an anti-ferrodistortive tetragonal structure. Interestingly, even at ultra-low temperature STO still does not have overall dipole moment, unlike many other perovskites which are ferroelectric at low temperatures. This is attributed to the quantum fluctuations over its shallow ferroelectric potential well~\cite{quantum_paraelectric_sto}. Above 105K~\cite{sto_105k_prb,sto_105k}, it exhibits a thermodynamically averaged paraelectric cubic \textit{Pm-3m} phase until its melting point at 2,300K~\cite{sto_2300k}. BaTiO$_3$ (BTO) is another typical material in the oxide peroskites family. It shows ferroelectric order below 400K~\cite{bto_curie_t} and 
it transforms into a cubic paraelectric phase above the Curie temperature. Same as many other perovskite materials, the cubic structure of BTO is an effect of averaging over tilted local configurations. This locally disordered nature has been shown in some puzzling observations in experiments, such as pyroelectricity and piezoelectricity~\cite{pyroelectric_in_bto}, 
second harmonic generation signals~\cite{bto_distort_shg}, and pair distribution function~\cite{pdf_in_bto} in paraelectric BTO. At the same time, this was also confirmed by many theoretical works~\cite{bto_afe_pnas,bto_distortion_xgZhao,perovskites_distortion_prb}.

\begin{center}
    \begin{figure}
        \centering
        \includegraphics[width=1\linewidth]{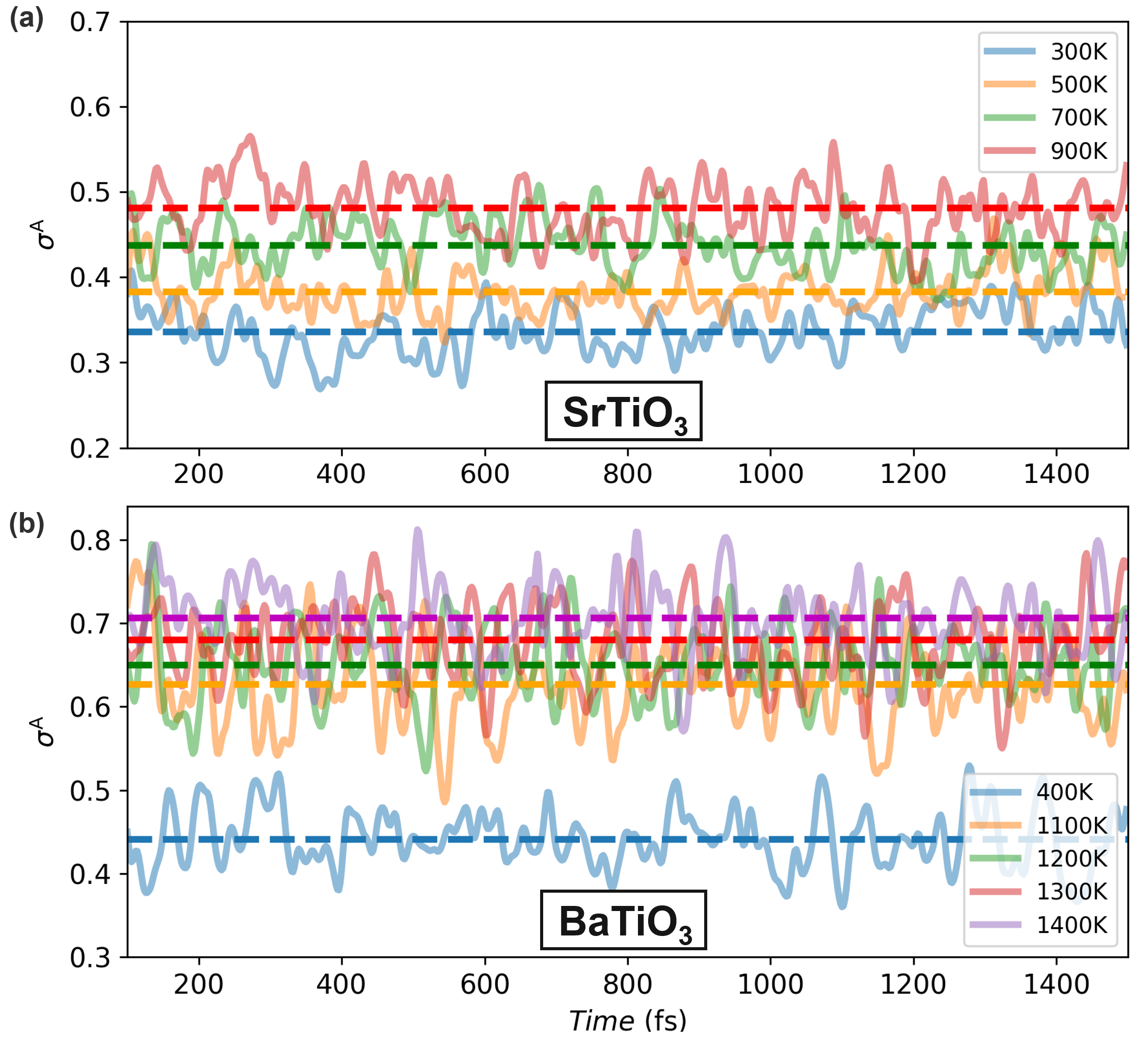}
        \caption{Time-dependent anharmonicity metric $\sigma^A$ of SrTiO$_3$ and BaTiO$_3$ at different temperatures as obtained via {\it ai}MD. The dashed lines indicate the mean value.}
        \label{fig:anharmonicity}
    \end{figure}
\end{center}

To quantify the anharmonicity of these materials, we use the anharmonicity metric~$\sigma^A$ proposed in Ref.~\cite{florain_prm}, which quantifies the difference between the actual anharmonic forces observed during {\it ai}MD and the approximative harmonic model. Accordingly, $\sigma^A$ vanishes for perfectly harmonic materials and nearly harmonic materials feature low $\sigma^A$~values,~e.g.,~$0.14$ and $0.08$ for silicon and diamond at 300K, respectively. As a rule of thumb, anharmonic effects become non-negligible in materials with $\sigma^A > 0.2$, which indicates that, on thermodynamic average, the anharmonic contributions make up for more than 20\% of the observed forces. Fig.~\ref{fig:anharmonicity}(a) shows $\sigma^A(t)$ for STO obtained from our 625-atom supercell {\it ai}MD simulation at different temperatures, whereby lattice thermal expansion
is accounted for in the simulations. Even at room temperature, SrTiO$_3$ exhibits already a high value of $\sigma^A = 0.34$; 
as the temperature increases, $\sigma^A$ also increases up to a value of 0.48 at 900K. At such high levels of anharmonicity, in which anharmonic effects contribute 
between 34\% and 48\% of the total forces, perturbative frameworks that treat these effect as a minor perturbation are no longer appropriate. For instance, 
this is reflected in the fact that the band gap renormalization of STO is severely underestimated when relying on the harmonic approximation, but can 
be reproduced when accounting for anharmonic effects~\cite{marios_bzu,marios_special_dispalcement}. The anharmonicity metric of BTO is shown in Fig.~\ref{fig:anharmonicity}(b). 
Just above the phase transition temperature at 400K, it already shows a very large anharmonicity $\sigma^A \approx 0.45$, comparable to STO at 700K. And as expected, the anharmonicity 
increases even further with rising temperature. At our highest studied temperature of $T=1400K$ it shows $\sigma^A \approx 0.70$, so anharmonic effects largely dominate the dynamics in this case.

\subsection{Electron Mobility of SrTiO$_3$}
\begin{center}
    \begin{figure}
        \centering
        \includegraphics[width=1\linewidth]{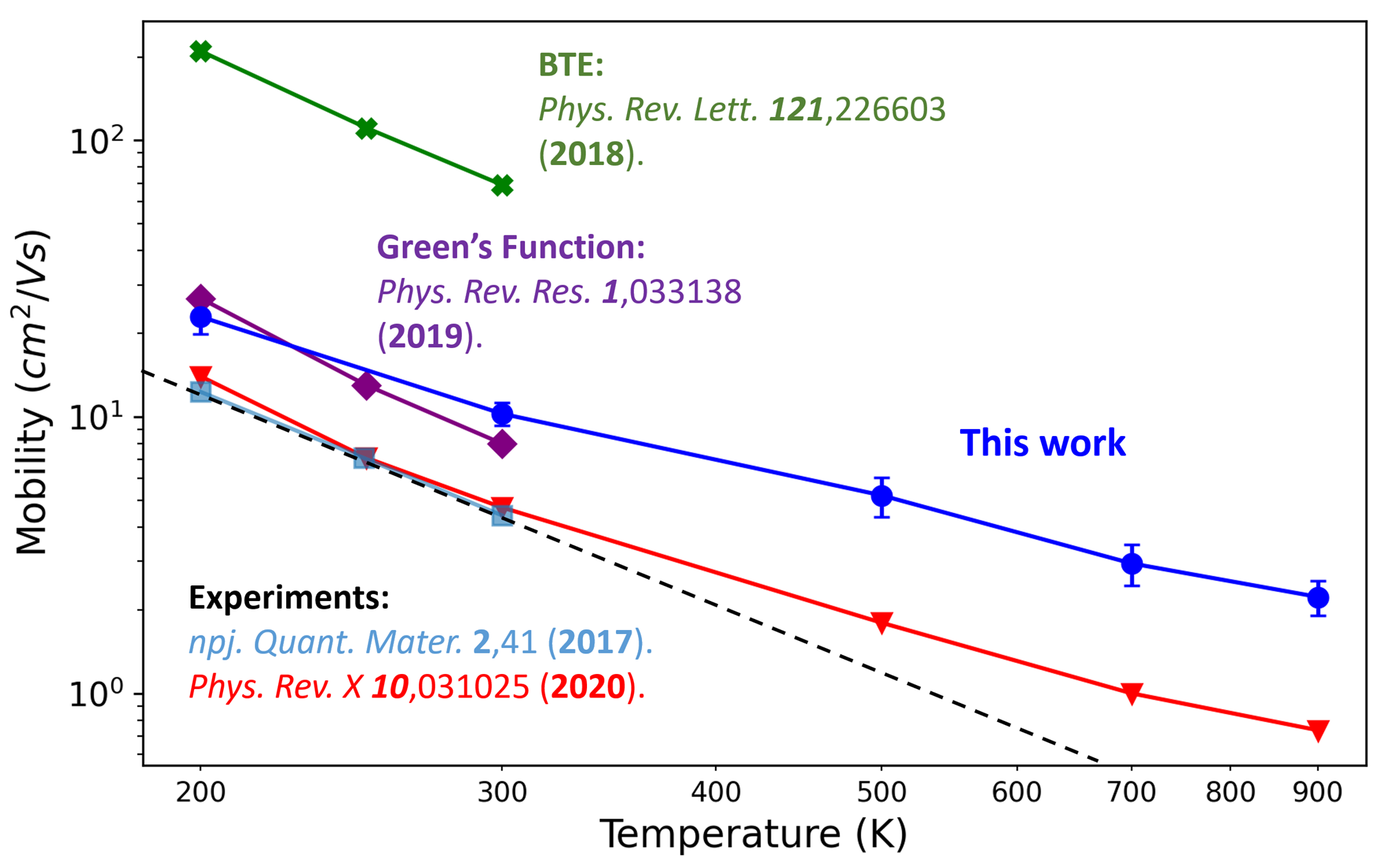}
        \caption{Temperature-dependent electron mobility of SrTiO$_3$ compared to BTE results~\cite{sto_bte}~(green line) and values obtained using the retarded Green's function approach~\cite{sto_spectral_function_method_prr}~(purple line). Experimental results (red~\cite{sto_experiment} data with carrier density $5.9\times 10^{18} cm^{-3}$, skyblue~\cite{sto_experiment_300k}) and the low-temperature $\sim T^{-3.1}$ trend~(black dashed line) are shown as well.}
        \label{fig:sto_result}
    \end{figure}
\end{center}

\begin{center}
    \begin{figure*}
        \centering
        \includegraphics[width=1\linewidth]{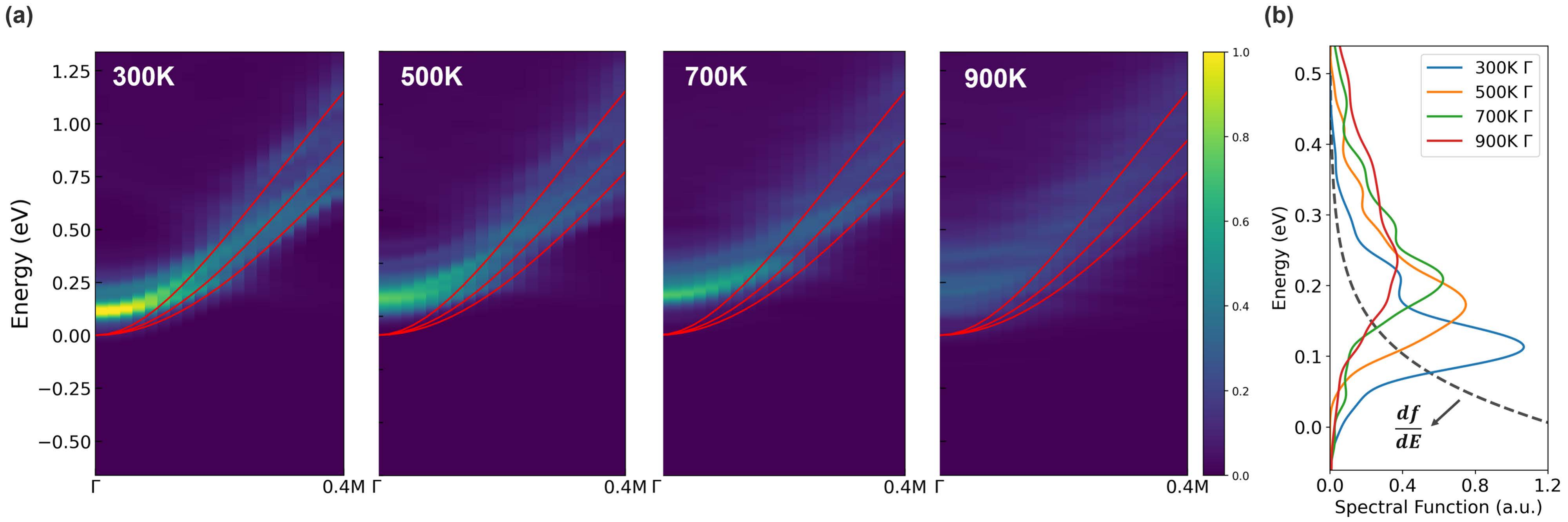}
        \caption{(a) Spectral function of SrTiO$_3$ near the conduction band minimum along the $\Gamma$ - $0.4$M path at different temperatures. The color bar indicates the spectral weight. The red curve show the band structure of the perfect primitive cell at 0K. The band gap renormalization and broadening of the spectral function with temperature is clearly observed. (b) Spectral function at $\Gamma$ point near conduction band minimum. As temperature increase, more and more spectral weight is transferred to the high energy tail. Black dashed line shows the derivative of the Fermi function.}
        \label{fig:sto_spectral_compare}
    \end{figure*}
\end{center}

As shown in Fig.~\ref{fig:sto_result} for SrTiO$_3$, our KG + {\it ai}MD calculation with a 625-atom supercell reproduces the correct, qualitative behaviour of the mobility up to very high 
temperature when compared to experiments, including the observed deviations from a low temperature $T^{-3.1}$ scaling at high temperature measured in Nb-dopped STO with $ < 10^{20} cm^{-3}$ carrier density~\cite{sto_experiment}.  In comparison, the state-of-the-art BTE methods can qualitatively capture the correct trend of the electron mobility 
up to room temperature (150-300K)~\cite{sto_bte}, but the value is one order of magnitude higher than the KG results, as shown by the green line in Fig.~\ref{fig:sto_result}. Note that these BTE calculations already
incorporate some anharmonic effects via temperature-dependent phonon dispersions. Including higher order electron-phonon interactions in the retarded Green's function can cover this gap and get quantitatively comparable 
results with the KG approach in the low temperature regime, as shown by the  purple line in Fig.~\ref{fig:sto_result}. 
However, in both cases, perturbation theory predicts a $\sim T^{-n}$ temperature dependence of the mobility~with~$n=-3.1$, as highlighted by the dashed line in Fig.~\ref{fig:sto_result}. 
Let us emphasize that perturbation theory is bound by construction to a constant $n$-value~\cite{ponce_review} for the whole high-temperature range and can thus not match 
the $\sim T^{-1.5}$ trend~\cite{sto_experiment} observed in experiment. At variance with that, the KG approach correctly captures changes in $n$ and also the $\sim T^{-1.5}$ behaviour at large temperatures. 

Quantitatively, we note that the calculated mobility values for the retarded Green's function and the KG approach agree at 300~K, but are both larger then the experimental results. We here assume that this discrepancy is caused
by impurity scattering in the doped samples. A simple estimation of impurity scattering by semi-empirical Brooks-Herring model as introduced in Sec.~\ref{sec: computational_details} gives total mobility $\sim 5.8 cm^2/Vs$ at 300K, which is closer to experimental value of $\sim 4.7 cm^2/Vs$.

Coming back to the  deviations from the $T^{-3.1}$ trend at high temperatures, our calculations can be rationalized using the fully anharmonic spectral functions 
shown in Fig.~\ref{fig:sto_spectral_compare}(a). With increasing temperature, the spectral function becomes broader and broader, whereby the width of the spectral 
function corresponds to the imaginary part of the electron self-energy. Conversely, the shifts observed for the peak maximum in the spectral function correspond 
to the real part of the self-energy~\cite{marios_bzu}. The massive broadening of the spectral function indicates very strong electron-vibration interactions at increasing 
temperature. For a closer analysis, we plot the temperature-dependent spectral function at the CBM,~i.e.,~the $\Gamma$-point, in Fig.~\ref{fig:sto_spectral_compare}(b). This reveals
that the spectral function becomes not only very broad, but also asymmetric at high temperatures, since it develops a long, high-energy tail. 
This tail does virtually not contribute to the mobility due to the fast decay of the Fermi occupation function, as shown by the derivative of 
Fermi function in Fig.~\ref{fig:sto_spectral_compare}(b). Thus, the mobility drops down with increasing temperature, since more and more spectral weight 
is transferred to the non-contributing tail. 
Let us emphasize that this asymmetric line-shape of the spectral function is a key signature for anharmonic and higher-order 
electron-vibrational coupling~\cite{thermal_conductivity_spectral_function_method}, that is not accounted for in state-of-the art perturbative BTE approaches. In such methods, the peaks of the spectral function are Lorentzian by construction~\cite{ponce_review}, so that the described high-temperature spectral weight shift remains unaccounted for. The asymmetricity of the spectral function of STO with temperature will be analyzed in details together with BTO in the next section.

\subsection{Electron Mobility of BaTiO$_3$}

In comparison to SrTiO$_3$,  BaTiO$_3$ exhibits a much lower mobility,~e.g., already 0.6 cm$^2$/Vs at 400K.
Also, the temperature dependence observed in experiments exhibits a quite different behaviour. As discussed above, SrTiO$_3$ follows a $T^{-n}$ law with $n=3.1$
at low temperatures. At more elevated temperatures, the decay becomes less steep to $n\sim 1.5$. Conversely, the decrease 
of mobility in BaTiO$_3$ show a $\sim T^{-1.24}$ behaviour over the whole temperature range. Note that some 
experiments argue that the mobility of BaTiO$_3$ should also level off and become constant at high temperatures~$T>1100$~K~\cite{bto_kim_0.16,bto_song_yellow_point,bto_yoo_0.13}. 
However, the scattering between different experiments, the error bars within one measurement~\cite{bto_song_yellow_point}, and the few available high-temperature data points
hardly allow to fully corroborate this conclusion from experiment alone.
 
\begin{center}
    \begin{figure}
        \centering
        \includegraphics[width=1\linewidth]{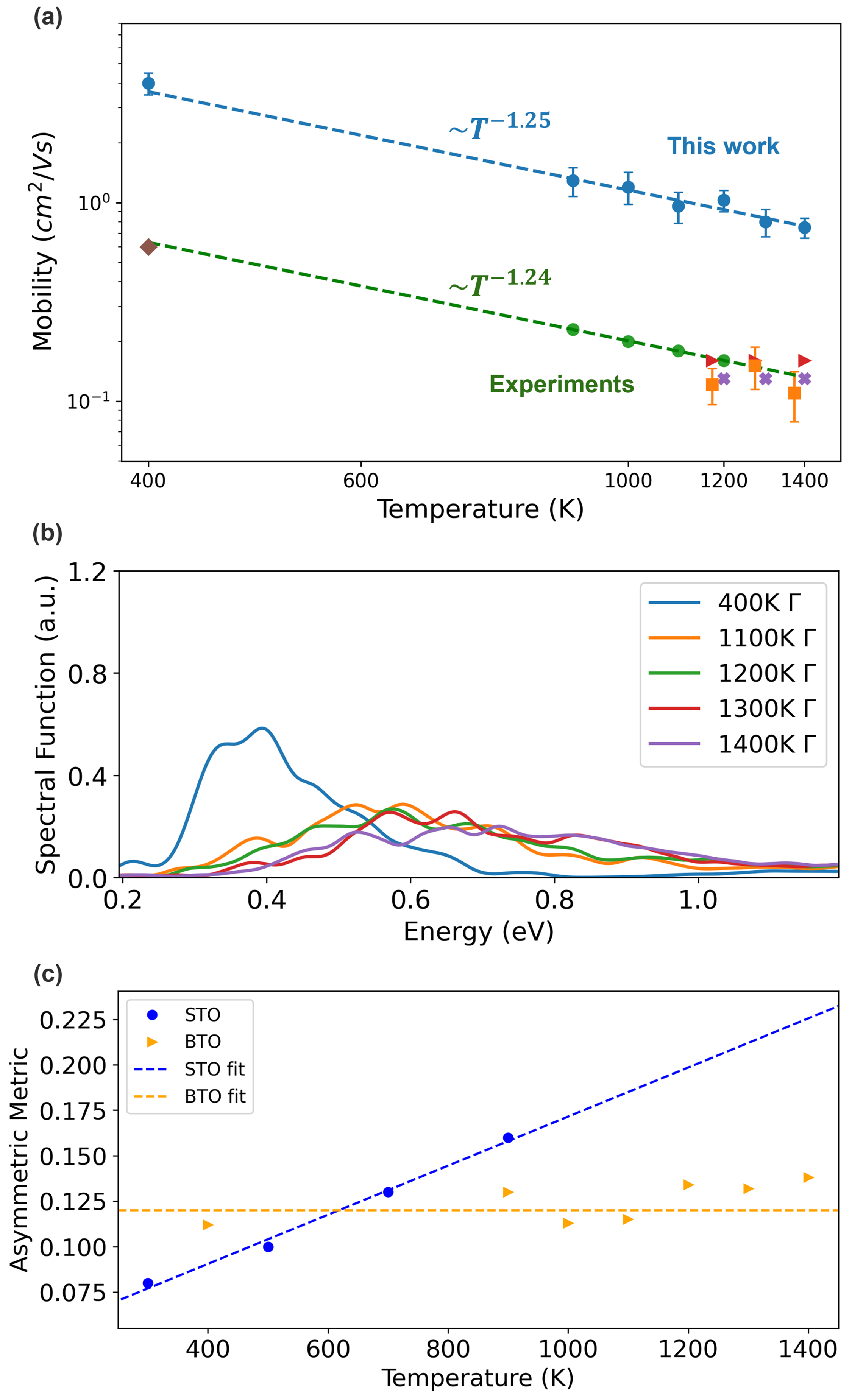}
        \caption{(a) Temperature dependent electron mobility of BTO. Blue dots are the result in this work. Others are experimental values: $\blacklozenge$~\cite{bto_mobility_400k_exp}, $\bullet$~\cite{bto_mobility_900_1200K}, $\blacktriangleright$~\cite{bto_kim_0.16}, $\blacksquare$~\cite{bto_song_yellow_point}, 
        $\times$~\cite{bto_yoo_0.13}.
        The dashed line indicates the $T^{-n}$ law.
        (b) Spectral function of BTO at the $\Gamma$-point near the conduction band minimum at different temperatures. 
        (c) Asymmetricity metric of STO and BTO.}
        \label{fig:bto_result}
    \end{figure}
\end{center}

The performed KG calculations for the electron mobility confirm that a $T^{-1.25}$ scaling law holds over a wide temperature region. 
{Strong anharmonicity can also be qualitatively illustrated in the spectral function of BTO at band edge. As shown in Fig.~\ref{fig:bto_result}(b), similar asymmetric spectral function is also observed. At high temperature (>1,100K), we observed that the spectral function looks similar, all show a very broad non-Lorentzian distribution, indicating the breakdown of the quasi-particle picture. The spectral peak height in BTO at 400K is lower than the spectral peak height of STO at 500K and comparable with that at 700K in Fig.~\ref{fig:sto_spectral_compare}(b). This trend is qualitatively related to the higher $\sigma^A$ of BTO at 400K compared to STO at 500K and is more similar to SrTiO$_3$ at 700K. The lower spectral peak in BTO than STO at fixed T can also explain the lower mobility in BTO.}

The different temperature dependent behaviour of STO and BTO can be qualitatively rationalized by the asymmetricity of their spectral functions. The method to evaluate asymmetricity of the spectral function is introduced in Sec.~\ref{sec: computational_details}. As shown in Fig.~\ref{fig:bto_result}(c), the asymmetricity of STO's spectral function increases with temperature. In comparison, the asymmetricity of BTO's spectral function barely changes for the whole temperature region. The nearly constant asymmetricity of the BTO spectral function indicates high similarity of the spectral shape with increasing temperature, which result in nearly constant $n$ in the $T^{-n}$ scaling law. On the other hand, the asymmetricity shows that the spectral function of STO  changes shape significantly with temperature, 
inducing changes in the $n$-scaling in the $T^{-n}$ law.

Eventually, we note that the predicted mobility values are higher than in experiment. 
This difference may comes from the fact that experiments actually use poly-crystalline~\cite{bto_song_yellow_point,bto_yoo_0.13} and highly-dopped BTO~\cite{bto_mobility_400k_exp} where grain boundaries and dopant can both reduce the measured mobility. Again, using the semi-empirical Brooks-Herring model, we can get an approximated total mobility of $\sim 1.4 cm^2/Vs$ at 400K, which is closer to experimental value of $\sim 0.6 cm^2/Vs$. 
Also some experiments argues that the formation of small polarons is important in BaTiO$_3$~\cite{polaron_review,bto_small_polaron_prb}. Such polarons can serve 
as additional scattering centers that are not accounted for in our simulation.

\section{Conclusion and outlook}
\label{sec:conclusion_kg}
In this work, we investigated and analysed the Kubo-Greenwood formula, a non-perturbative, first-principles approach for computing electronic transport based on fully anharmonic {\it ai}MD
calculations that is notoriously tricky and costly to converge for ordered, crystalline materials. To this end, we implemented the KG-formula and several numerical strategies that 
alleviate these computational hurdles in the numeric atom-centered orbital basis code \texttt{FHI-aims}. This includes Fourier interpolation to speed-up BZ integrations, a workflow to systematically choose the 
optimal combination of $\vec{k}$-grid density and broadening parameter in the KG formula, a workflow to achieve convergence with respect the thermodynamic phase-space sampling 
both with respect to number of samples and supercell size, and, eventually, a scheme to obtain correct mobilities from semi-local DFT calculations despite the well-known failure 
of such approaches in predicting band-gaps and charge carrier densities. Eventually, we show that the combination of these techniques enables to compute converged mobilities from
first principles, even for ordered crystalline solids that were numerically not accessible before. Compared to existing pertubative approaches~\cite{ponce_review}, the KG formalism has
the unique advantage that anharmonic effects in the nuclear dynamics and higher-order couplings between electronic and nuclear degrees of freedoms are naturally included.

We demonstrated the merits of the implemented KG formalism by computing the mobilities of the oxide perovskites SrTiO$_3$ and BaTiO$_3$ over a wide range of temperatures, reaching 
good agreement with experimental data. For instance, state-of-the-art BTE approaches for SrTiO$_3$ overestimate the mobility at room temperature by an order of magnitude,
whereas the KG formalism yields results closer to experiment, in line with retarded Green's function approaches~\cite{sto_spectral_function_method_prr}. At variance with this approaches, the KG formalism is 
furthermore able to correctly reproduce the experimentally measured temperature dependence of the mobility at higher temperatures, at which anharmonic and higher-order coupling 
effects become more prevalent and the quasi-particle approximation breaks down.  For SrTiO$_3$, we find that the mobility follows a $\sim T^{-3.1}$ scaling in the low temperature regime,
but exhibits a much less steep behavior of $\sim T^{-1.5}$ when the temperature increases, in line with high-temperature experiments~\cite{sto_experiment}. Conversely, BaTiO$_3$ 
shows a $\sim T^{-1.25}$ temeprature dependence across a wide temperature regime. We can rationalize this contrasting trend by analyzing the temperature-dependent spectral functions 
of these materials at various temperatures. For both materials, the electronic states relevant for the mobility
exhibit asymmetric, non-Lorentzian lineshapes. For SrTiO$_3$, the asymmetricity increases with temperature, so that the spectral weight of the conductive states at the bottom of the band
decreases. We ascribe the observed changes in the $T^{-n}$ scaling law to this effect. Accordingly, the asymmetricity of the spectral function of BaTiO$_3$  barely changes with temperature 
and also the scaling exponent~$n$ remains constant over a wide temperature range. Let us emphasize that this is a pure higher-order effects, since low-order perturbative 
approaches are limited to Lorentzian lineshape by construction.

In a nutshell, our work demonstrates the importance of anharmonic and higher-order coupling effects for the computation of mobilities and describes strategies on how to systematically account for
them using the KG formalism in first-principles approaches. With that, the described implementation can serve as a benchmark method to validate approximations taken in perturbative approaches. 
To this end, machine-learning based approaches appear desirable to further accelerate such KG calculations,~e.g.,~by avoiding explicit {\it ai}MD simulations and using machine-learned 
interatomic potentials for the dynamics instead~\cite{nequip,mlp_review_behler_gabor}. Similarly, machine-learning approaches for predicting electronic properties such as densities and
matrix elements~\cite{deeph,ml_matrix_james_kermode,deeph_dfpt,salted} provide a route to also accelerate the evaluation of the KG formula. 
{However, we also note the challenge that much of the anharmonicity of atomic motions are triggered by intrinsic defects explored in {\it ai}MD~\cite{florian_gk_application} that are typically not well captured by state-of-the-art machine learning potentials~\cite{kang2024active}.}

All the electronic structure theory calculations produced in this work and the {\it ab initio} molecular dynamics trajectories are available on the Novel Materials Discovery (NOMAD) repository under DOI~({DOI added upon acceptance.}).

\begin{acknowledgments}
This project was supported by the ERC Advanced Grant TEC1p (European Research Council, Grant Agreement No. 740233.)
J.Q. acknowledges support from Professor Angel Rubio and the Max-Planck Graduate Center for Quantum Materials (MPGC-QM).
M.S. and J.Q. acknowledge perceptive discussions with Zhenkun Yuan on basic aspects and several details
of this work.
J.Q. would like to thank Mariana Rossi, Shuo Zhao and Florian Fiebig for fruitful discussions.
\newline
\end{acknowledgments}

\begin{appendix}
\section{Merit of the Fourier interpolation}
\label{appendix_fi}
\begin{center}
    \begin{figure}
        \centering
        \includegraphics[width=1\linewidth]{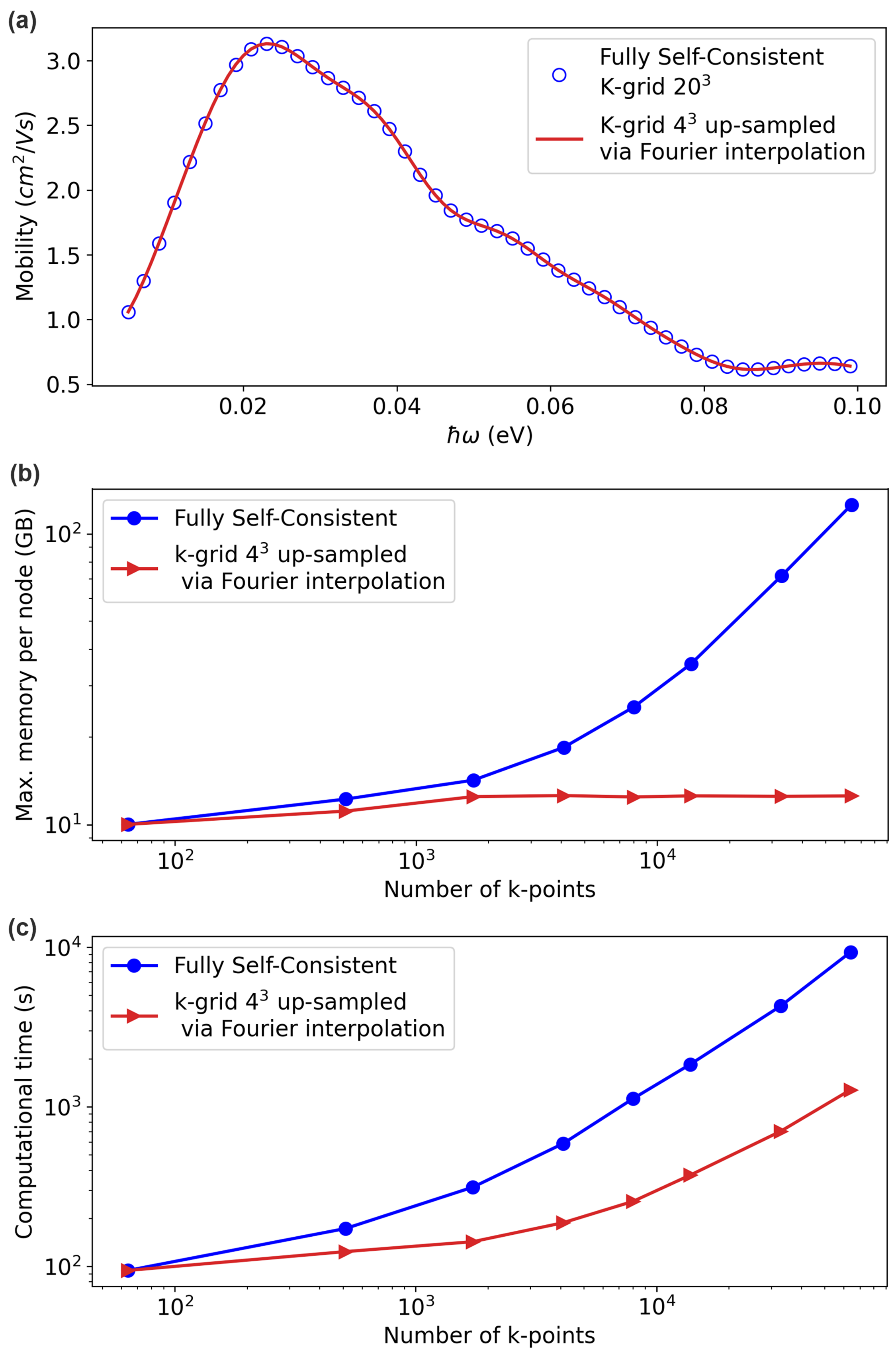}
        \caption{Merits of Fourier interpolation. (a) Comparison of the spectrum calculated from a dense $20\times 20\times 20$ SCF k-grid and a sparse $4\times 4\times 4$ SCF k-grid then do fourier extrapolate. (b) Comparison of maximum memory per node with increasing number of k-points. (c) Comparison of computational time with increasing number of k-points.}
        \label{fig:fourier_inter}
    \end{figure}
\end{center}

Fourier interpolation is a well-established technique in codes using linear combinations of atom-centered orbitals,~e.g.,~for band structure and density of state calculations.
The merits of this approach are shown in 
Fig.~\ref{fig:fourier_inter} for a representative 40 atoms,  $2\times 2\times 2$ SrTiO$_3$ supercell from an {\it ai}MD trajectory at 500K with $n_e = 1\times10^{18} (1/cm^3)$.
Thereby, we compare a fully self-consistent evaluation of the frequency-dependent mobility with one obtained via Fourier interpolation. In the first
case, a SCF \textbf{k}-grid $20^3$ points per cell is used for both the SCF cycle and the Kubo-Greenwood formula. In the second case, only a sparse \textbf{k}-grid $4^3$ is 
employed during SCF; for the evaluation of the Kubo-Greenwood formula the \textbf{k}-grid is then up-sampled to $20^3$ \textbf{k}-points via Fourier interpolation. As evident
from Fig.~\ref{fig:fourier_inter}(a) the resulting mobilities are virtually indistinguishable in the two cases. With respect to memory consumption and computational cost, the Fourier interpolation
is however extremely advantageous. The memory consumption is significantly lowered, since only the {\bf k}-dependent KS wave functions included in the SCF need to be actually stored concurrently.
For the evaluation of the  Kubo-Greenwood formula, the required KS wave functions are computed consecutively, so that the memory consumption remains constant, as shown in Fig.~\ref{fig:fourier_inter}(b). 
Since converging the mobility often  requires 10$^4$ $\vec{k}$-points and more, this results in memory savings of one order of magnitude and more. 

Similarly, we observe notable savings in computational time, as shown in Fig.~\ref{fig:fourier_inter}(c). In the case of a fully self-consistent solution, one diagonalization is required for each $\vec{k}$-point
in {\bf each} SCF step, whereas each of the Fourier-interpolated $\vec{k}$-points requires only {\bf one} diagonalization. Overall, this leads to almost an order of magnitudes in savings; 
the overall scaling remains similar, though, since the diagonalization itself is the computationally dominant step. 
 In this context, it is worth mentioning that the here employed Fourier interpolation is also the basis for the Wannier interpolation commonly employed for plane-wave basis 
sets~\cite{giustino_wannier_interpolation}. In that case, a Wannierization procedure is used to obtain a localized representation in real space that can be Fourier interpolated. For the case of NAO basis 
sets used in {\tt FHI-aims}, this first step is unnecessary since the representation is localized by construction.

\end{appendix}


%

\end{document}